\pdfoutput=1
\documentclass[preprint2]{aastex}
\usepackage{xfrac}
\usepackage[caption=false]{subfig}
\usepackage{hyperref} 
\hypersetup{
	breaklinks,
	colorlinks=true,
	pdfauthor={John Arban Lewis III},
	pdftitle={Protostars at Low Extinction in Orion A},
	citecolor=blue
	}
\usepackage{longtable}
\usepackage{color}

\newcommand{\ppcell}[1]{\includegraphics[width=2in]{#1}}
\newcommand{\source}[1]{\plottwo{#1a}{#1b}}
\newcommand{\M}[1]{M#1}
\newcommand{\msun}{M_\odot}
\newcommand{\kms}{{\rm~km~s^{-1}}}
\newcommand{\pc}{{\rm~pc}}
\newcommand{\ak}{A_{\rm K}}
\newcommand{\av}{A_{\rm V}}
\newcommand{\akv}[2]{#1 (#2)}
\newcommand{\samplesize}{44} % Number of protostars at low extinction
\newcommand{\stageI}{10}     % Number classified as Stage I
\newcommand{\pstageI}{23}
\newcommand{\stageII}{18}    % Number classified as Stage II
\newcommand{\pstageII}{41}
\newcommand{\gal}{4}         % Number of suspected galaxies
\newcommand{\pgal}{9}
\newcommand{\unk}{3}         % Number of unknown "real" sources  (TLDs)
\newcommand{\punk}{7}
\newcommand{\fuzz}{9}        % Number of fuzz + cloud edges
\newcommand{\pfuz}{20}
\newcommand{\tld}{11}        % Number of "unknown real" and "fuzz"

\begin{document}
%====Setup document title and author
\shorttitle{Protostars}
\shortauthors{Lewis \& Lada}
\title{Protostars at Low Extinction in Orion A}
\author{John Arban Lewis \& Charles J. Lada}
\affil{Harvard-Smithsonian Center for Astrophysics \\ 60 Garden St \\Cambridge, MA 02138}
\email{jalewis@cfa.harvard.edu} %corresponding author
\email{clada@cfa.harvard.edu}
\accepted{05/07/2016}

%\slugcomment{\color{red}\bf DRAFT - DO NOT CIRCULATE}

\begin{abstract}
In the list of young stellar objects compiled by \citet{megeath2012} for the Orion A molecular cloud, only \samplesize\ out of 1208 sources found projected onto low extinction ($\ak<0.8$ mag) gas are identified as protostars. These objects are puzzling because protostars are not typically expected to be associated with extended low extinction material. Here, we use high resolution extinction maps generated from Herschel data, optical/infrared and Spitzer Space Telescope photometry and spectroscopy of the low extinction protostellar candidate sources to determine if they are likely true protostellar sources or contaminants. Out of \samplesize\ candidate objects, we determine that \stageI\ sources are likely protostars, with the rest being more evolved young stellar objects (\stageII), galaxies (\gal), false detections of nebulosity and cloud edges (\fuzz), or real sources for which more data are required to ascertain their nature (\unk). We find none of the confirmed protostars to be associated with recognizable dense cores and we briefly discuss possible origins for these orphaned objects. 
\end{abstract}

\keywords{extinction,galaxies:star formation,infrared:general,stars:protostars}

\maketitle

\section{Introduction}\label{intro}

Accurate characterization of young stellar objects (YSOs) is important for studies of star formation. For example, comparing the relative numbers of protostars and more evolved pre-main sequence (PMS) stars can provide information regarding the relative lifetimes of the various stages of early stellar evolution. Comparison of the relative positions of YSOs in different evolutionary states with the location of dense gas structures can also yield information regarding the diffusion of YSOs in a star forming cloud. A reliable census of YSOs is also necessary for determining accurate star formation rates (SFRs) and efficiencies (SFEs) in molecular gas. In particular, because of a protostar's extreme youth, a dependable census of a cloud's protostellar population is essential for constructing important empirical relationships, such as the Schmidt Law, that probe the physical connection between star formation and molecular cloud structure \citep[e.g.,][]{2013ApJ...778..133L}.
 
Recent studies of nearby molecular clouds suggest the existence of a column density threshold for star formation, below which stars rarely form. The estimated value of the threshold is $\ak \sim 0.8$ magnitudes and has been determined by comparing maps of a cloud's extinction with the numbers and/or locations of protostars within the cloud \citep[e.g.,][]{2004ApJ...611L..45J,2010ApJ...723.1019H,2010ApJ...724..687L}. However, many of these same molecular clouds also exhibit Schmidt-like (i.e. power-law) scaling relations between the surface densities of YSOs and cloudy material over a wide range of extinctions with no evidence for a break in the relation near the suggested threshold \citep[e.g.,][]{guter2011, 2013ApJ...778..133L, 2014A&A...566A..45L}. This apparent paradoxical situation is largely resolved by the fact that the derived power-law slopes of the Schmidt relations are found to be relatively steep indicating extremely low protostellar surface densities in the low extinction regions compared to the high extinction regions of molecular clouds. Indeed, observations find that only about 10\% of all protostars are projected on regions below the proposed extinction thresholds \citep{2013ApJ...778..133L, 2014ApJ...782..114E}. Despite their relative rarity, the presence of protostars in low extinction regions is of considerable interest, however, since it raises interesting questions about the very nature of protostars, in particular, whether protostars can form {\sl in situ} in such rarified conditions.

In order to understand the nature of low extinction protostars one must first reliably verify their existence. YSOs are best identified and classified by the shapes of their spectral energy distributions (SEDs). The shapes of YSO SEDs, particularly in the infrared, are sensitive to the presence, structure and amount of surrounding circumstellar dust and are therefore indicative of the evolutionary state of the object \citep{lw1984, 1987IAUS..115....1L, 1987ApJ...312..788A}. SED classes (i.e., 0, I, II and III) describe a YSO sequence characterized by decreasing amounts of circumstellar dust and generally correspond to three basic YSO evolutionary states or stages \citep{robit2006}: protostar (Stage I), pre-main sequence (PMS) star with circumstellar disk (Stage II), and dust free or naked PMS star (Stage III). Protostars correspond to SED classes 0 and I, depending, respectively, on whether there is substantially more mass in the envelope than in the central star and disk \citep{1993ApJ...406..122A} or vice versa. Protostellar evolution is driven by the depletion (or removal) of the surrounding circumstellar envelope. Once the envelope is dissipated, a circumstellar disk often remains and dominates the infrared emission from the YSO resulting in a Class II SED. Given wavelength complete SEDs, protostars (Class 0$+$I sources) can be readily distinguished from Class II or disk dominated sources. Unfortunately, it is not practical to obtain complete SEDs for statistically significant samples of YSOs in molecular clouds. In such circumstances YSOs are typically identified by the presence of an infrared excess and classified by color indices that measure the slope of the SED over some limited wavelength range. One problem with such classifications is that degeneracies can arise between protostars and disk dominated (Class II) sources due to unrelated intra-source extinction \citep[e.g.,][]{2009A&A...498..167V,forbrich2010} and even disk inclination \citep[e.g.,][]{1997ApJ...490..368C}. Although the former issue is not likely to be of concern in extended regions of low extinction, the latter issue could lead to significant misclassifications of Stage II sources as protostars \citep{2008A&A...486..245C} in such regions. Degeneracies can also arise between YSOs and background contaminants such as carbon stars, AGB stars and galaxies; and this effect would be more pronounced in the low extinction regions that are of interest for this paper because such contaminants would not be as easily screened out as in high extinction regions. Given these concerns, a closer examination of low extinction protostars seems warranted.

\begin{figure*}[htb]
\centering
\includegraphics[width=5in]{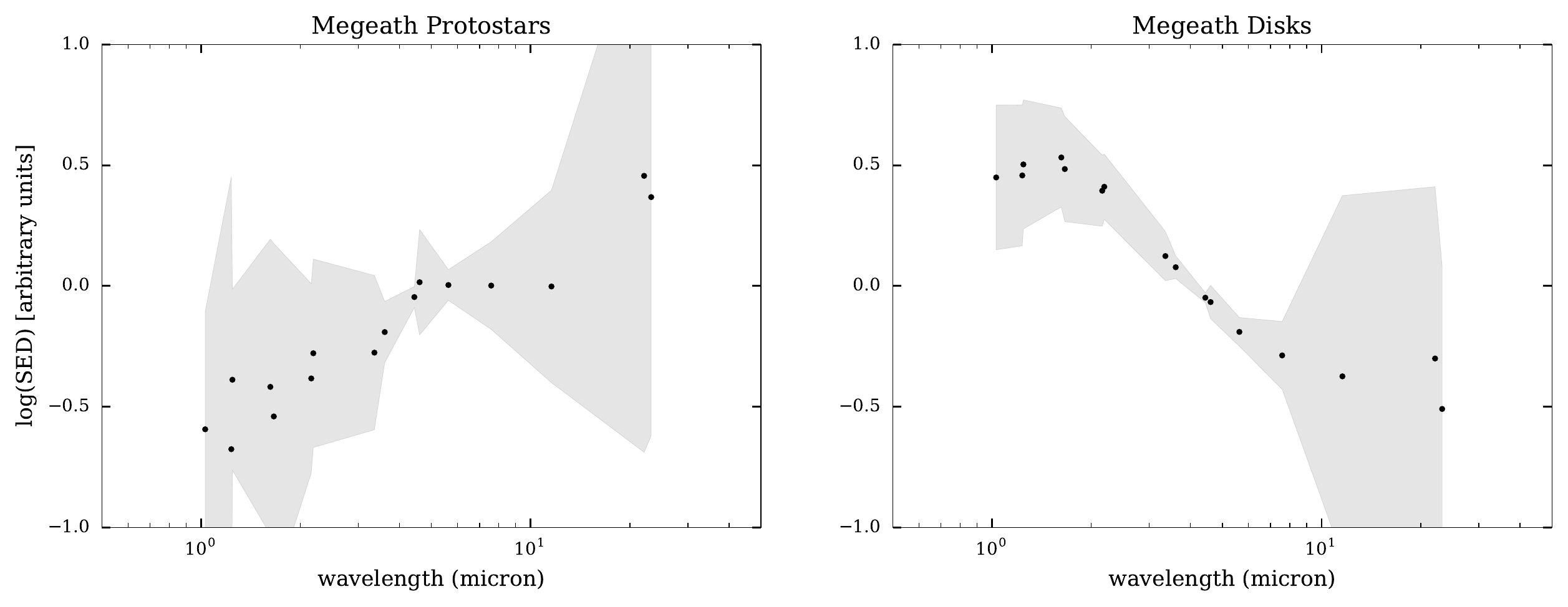}
\caption{\footnotesize\label{fig:med_all} Median SEDs of Class 0/I protostellar sources (left) and of disk dominated Class II sources (right) as classified by \protect\citet{megeath2012}. The SEDs are normalized by the average IRAC Band 1-3 Flux. The grey shaded regions represent the 1 $\sigma$ dispersion of the data in each panel. The slopes of the median SEDs measured between $2-8$ microns are $+0.75\pm0.18$ and $-1.33\pm0.07$, respectively. Note that the latter slope is the same to that (-1.33) expected for a flat accretion or reprocessing disk. }
\end{figure*}

Perhaps the best region for the investigation of low extinction protostars is the Orion A molecular cloud. Using Spitzer observations, \citet{megeath2012} conducted a census of YSOs in Orion and classified 329 infrared excess sources as protostars using infrared colors and brightnesses. In general, their color-based method appears to be able to separate protostars from disk dominated YSOs. This is illustrated in Figure \ref{fig:med_all} which shows that the median SED shape of the Megeath et al. protostars is distinct from that of YSOs they classified as disk dominated objects. \cite{2013A&A...559A..90L} used the Megeath catalog, along with extinction maps derived from the 2MASS survey using the NICEST method \citep{2011A&A...535A..16L}, to determine a Schmidt-like scaling law, $\Sigma _{*}\approx \kappa A_K^\beta$ (power-law index $\beta$, normalization $\kappa$), relating the surface density of protostars to that of the dust in Orion A. They found $\beta$ = 2.0 indicating a steep non-linear rise of protostellar surface density with extinction. These observations showed that 44 objects, or roughly 13\% of the protostellar population identified by \citet{megeath2012}, were located in extended regions of relatively low extinction, below the nominal star formation threshold of $\ak$ $\approx$ 0.8 magnitudes and 27 objects (or 8\%) were located in regions with extinctions $<$ 0.5 magnitudes. In these regions there are roughly 10$^3$ objects classified as Class II sources by \citet{megeath2012} and misclassification of only few \% of the Stage II population as Class I objects might account for all the low extinction protostars identified in these low extinction regions. Indeed, analysis of SED models by \citet{2008A&A...486..245C} suggests much higher fractions of Stage II sources can be routinely misclassified as Class I objects. 

In this paper, we re-examine the classification of low extinction protostars from the \citeauthor{megeath2012} catalog using stringent criteria based on SED modeling and informed by spectroscopic and imaging data from the literature.  We identify sources that are most likely to be protostars and consider the question of whether they formed in situ at low extinction. In \S\ref{method}, we discuss the sample of low extinction protostars and our data selection criteria for building their SEDs. We also describe our classification criteria. In \S\ref{results}, we present the results of running our sample through our classification scheme and discuss interesting objects and contamination. In \S\ref{discuss}, we discuss three alternate schemes for explaining the protostars presence at low extinction and describe our conclusions in \S\ref{conclus}.

\section{Methodology}\label{method}

\subsection{Source Selection\label{selection}}

\begin{figure*}[ht]
\centering
\includegraphics[width=6.5in]{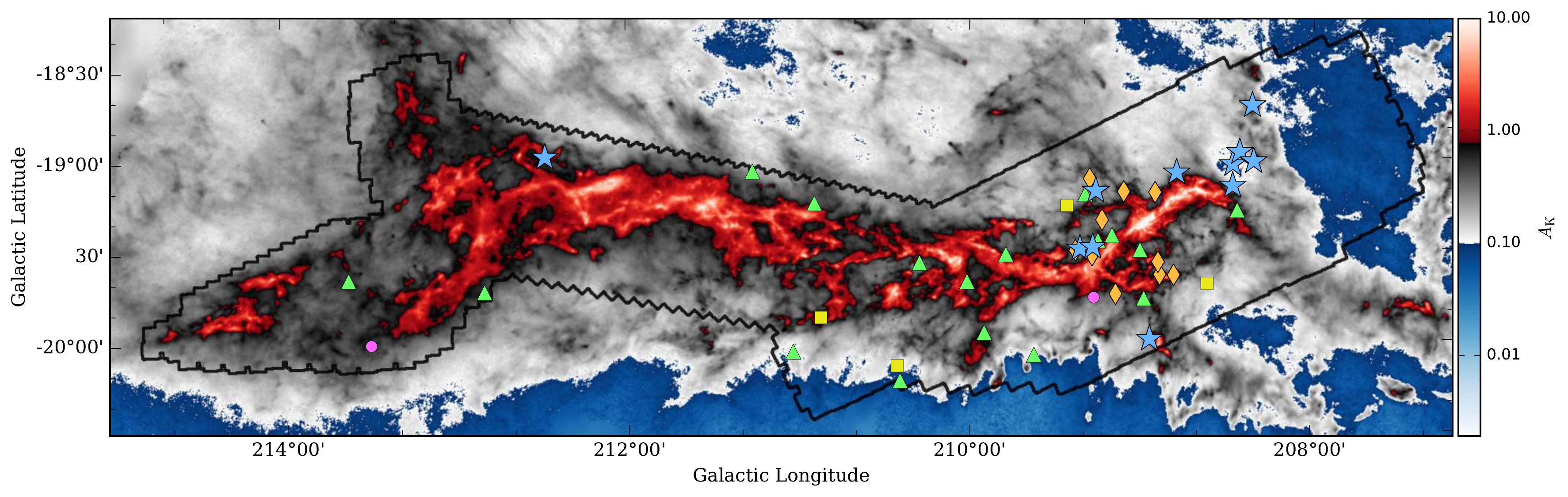}
\caption{\footnotesize \label{fig:all}{\it Herschel} K-band extinction ($\ak$) map of Orion A molecular cloud (from Lombardi et al. 2015) with low extinction protostellar candidates from \protect\citet{megeath2012} overlaid. The black line represents the boundary of Megeath et al. survey region. For the color scale of the extinction map blue corresponds to $\ak < 0.1$, gray to $0.1<\ak<0.8$, and red to $\ak>0.8$. The revised source classifications of this paper are represented by blue stars (confirmed protostars), green triangles (disk dominated sources), yellow squares (galaxies), orange diamonds (fuzz, cloud edges), and magenta circles (unclassified point-like sources).  }
\end{figure*}

We use the YSO catalog of \citet{megeath2012} and the extinction maps from \citet{2014A&A...566A..45L} to identify the protostellar candidates that are projected on low (i.e., $\ak<0.8$ mag) extinction regions of the Orion A cloud. \citet{megeath2012} used Spitzer and 2MASS color-color and color-magnitude diagrams to identify and classify 2818 YSOs in Orion A, of which 329 were classified as candidate protostars. The {\it Herschel}-derived extinction maps of Orion have an angular resolution of $36\arcsec$, and are sampled $15\arcsec$ intervals. At the distance to Orion (420 pc), 15$\arcsec$ corresponds to approximately 0.03$\pc$.  For our sample we selected those candidate protostars whose positions were coincident with extinction map pixels characterized by $\ak<0.8$ mag. Of the 329 candidate protostars in Orion A, only 44 are projected on low extinction (i.e., $\ak<0.8$ mag) pixels. We refer to these sources by their index in the Megeath catalog (M\#), which is a shorthand for the SIMBAD and virtual observatory compatible nomenclature MGM2012 \#. Figure \ref{fig:all} shows the extinction map for the Orion A cloud with the locations of the protostars identified at low extinction.

\subsection{Source SEDs\label{sourceseds}}

To further refine the \citeauthor{megeath2012} YSO classifications, we compile and examine SEDs (expressed as $\lambda F_\lambda$ vs. $\lambda$) for all 44 low extinction protostars in Orion A and for similarly sized samples of candidate protostars in the high extinction regions of the cloud. We then fit the SEDs with models generated by \citet{robit2006} to re-classify the sources and provide more secure classifications of these objects.

\subsubsection{Photometry\label{photometry}}

 SEDs generated from \citeauthor{megeath2012} are not complete in all bands for all sources. Incomplete SEDs make accurate source classification difficult, if not impossible. In order to improve the wavelength coverage of the SEDs, we use the \texttt{Vizier} catalog access tool to add data from WISE \citep{2010AJ....140.1868W}, the UKIDSS Large Area Survey \citep{2007MNRAS.379.1599L}, and SDSS \citep{2012ApJS..203...21A}. In each catalog we searched for the nearest source with clean photometry that was within $1.2\arcsec$ of the Megeath position. In SDSS we required the photometry to be flagged as \texttt{clean} (\texttt{q\_mode=1}) and magnitudes to be above the SDSS zero-flux limit. Any photometry reported with 0 error was treated as bad data and thrown out. In UKIDSS we only selected primary (\texttt{mode=1}) photometry for sources whose profiles were well modeled by a point spread function ({\texttt{cl=-1|-2}). For WISE we accept the matched photometry from the ALLWISE Data Release with no further filtering or refinement. We identified one close binary (separation $< 2\arcsec$), \M845, in SDSS. It is the only source which we resolve into a close binary. We use only the component closest to the original Megeath et al. coordinates and note that M845 is blended in all other bands. Finally, we incorporate recently published {\it Herschel} PACS 70$\micron$ photometry  for 9/44 sources in our sample observed in the Herschel Orion Protostar Survey (HOPS) \citep[][preprint]{furlan2016}. Using the data available in the preprint, we estimate the photometric error to be $\sim$5\%. 

 Of the \samplesize\ candidate low extinction protostars \tld\ had three or fewer data points and consequently SEDs that were not sufficiently complete to enable reliable classification. Upon close examination of the various images, \fuzz\ out of \tld\ of these sources were determined to be false positive identifications, that is they were mostly extended fuzzy, non-stellar cloud structures or edges and were removed from further analysis. The two remaining sources, M0122 (detected in J, 8\micron, and 24\micron) and M1203 (detected in 8\micron\ and 24\micron), were stellar-like and are assigned an "unknown" classification.

\subsubsection{IRS Spectra\label{irsinit}}
We augment the photometric SEDs with publicly available IRS spectra from multiple programs downloaded from the NASA Infrared Science Archive (IRSA). The data is available reduced and calibrated. Calibration errors are evident in some spectra due to offsets between the IRS spectral bandpasses. We do not correct for these since we are only interested in the existence or non-existence of silicate (Si) emission/absorption lines at 9.7$\micron$. Because of the lack of significant foreground extinction, silicate absorption observed toward the low extinction protostellar candidates is likely indicative of an optically thick protostellar envelope. Silicate emission arises from material near the photosphere of a YSO, likely within a circumstellar disk and it is usually only detectable because any intervening material is optically thin \citep[see][]{2008A&A...486..245C}. The Si feature is discussed more in \S\ref{sires}.

\subsection{SED Model Fits and YSO Classification\label{sedmodeling}}

\subsubsection{SED Fitting Procedure\label{sedfitproc}}

To refine the color-based classifications of \citeauthor{megeath2012}, we employ model fits to the source SEDs to derive certain physical parameters (described below) that enable us to assess the relative importance of circumstellar envelopes and disks around each object. We fit the source SEDs using the library of models from \citet{robit2006} which are based on the YSO model developed by \cite{whitney2003}. This library consists of $2 \times 10^5$ models computed by first assuming a given structure for the envelope and disk and the properties of dust and then by varying 14 physical parameters to produce a set of 20,000 physical models sampled at 10 different viewing angles. The 14 parameters that are varied over a large, astrophysically motivated, range of values include the accretion rate, radius, and opening angle of the envelope, the inner and outer radii, scale height, and flaring of the disk, and the mass, radius, and temperature of the central star, etc. We employ the \texttt{sedfitter}\footnote{\url{https://github.com/astrofrog/sedfitter}} Python module based on \citet{robit2007} to perform the fits. For fitting low extinction sources we allowed the foreground extinction to vary from zero to the extinction value of the {\it Herschel} pixel for every source and the distance to vary between 390-450 pc. The distance range is based on the various distances for Orion quoted in the literature \citep[e.g.,][]{2007A&A...474..515M, 2014ApJ...786...29S}. We adopt a distance of 420 pc for calculations done in this paper. The fitter fits all $2\times10^5$ models within these ranges of extinctions and log-spaced distances and selects the $(A_{\rm V}, \log{d_{kpc}})$ pair with the lowest $\chi^2$ for each model.

We use those fits with $$\Delta\chi^2_{scaled} \equiv \sfrac{\left(\chi^2 - \chi^2_{best}\right)}{\chi^2_{best}} \le 1$$ to select the models which are consistent at the $\sim\!2 \sigma$ level with the data. $\Delta\chi^2_{scaled}$ effectively scales up the errors by a constant value so that the lowest reduced $\chi^2$ is 1. We prefer this to setting an arbitrary fractional error or error floor for each band since the relative precision of each band varies depending on the source SED and the background where it is located. We use the $\chi^2$ weighted ($w_i = 1/\chi_i^2$) median of the fit parameters to determine the values of the physical parameters (i.e., stellar mass, envelope mass, infall rate) that are used to classify a source SED as protostellar. 

\subsubsection{\label{protcrit}Protostellar Classification Criteria}

We use the concept of evolutionary stages, developed in \citet{robit2007}, to construct a definition of what we consider a protostar for the purposes of this study. We use slightly different definitions for the stages than were introduced in \cite{robit2007}. Stage I sources are true protostars, with envelope dominated (Class I) SEDs and active infall. Stage II YSOs have have disk dominated SEDs and only a weak or non-existent envelope. Stage III sources are PMS stars with stellar dominated (Class III) SEDs and no circumstellar material. For a protostar we require that the ratio of envelope mass to stellar mass to equal or exceed 0.05, and that the ratio of mass infall rate to stellar mass be equal to or exceed 10$^{-6}$ yr$^{-1}$, that is,
\begin{description}\itemsep3pt \parskip0pt \parsep0pt
\item[Stage I:] $\dot{M}_{env}/M_{\star} \ge 10^{-6}\, {\rm yr^{-1}}$,\\ $M_{env}/M_{\star} > 0.05 \msun$
\item[Stage II/III:] $\dot{M}_{env}/M_\star<10^{-6}~{\rm yr}^{-1}$,\\ $M_{env}/M_{\star} < 0.05 \msun$
\end{description}
At face value these are very conservative criteria, for example protostars surrounded by such small envelopes would be in a very late stage of evolution, having accreted 95\% of their final masses. The implied limit on the evolutionary timescale of $<$ 10$^6$ yrs is certainly consistent with but on the upper end of current estimates of protostellar lifetimes (2.5-5 $\times$ 10$^5$ yrs.). 

In order to obtain final classifications for the low extinction protostellar candidates we augment the SED fitting results with Spitzer IRS data where available. We derive the final classifications for the low extinction protostellar candidates in Orion A using the following criteria:
1) Sources with IRS spectra are examined for the presence of a silicate emission or absorption feature. We assign a Stage II classification to the sources with Si emission, except in the case that the SED models indicate the source to be a face-on protostellar object, in which case a classification as a protostar is assigned.
2) The remaining sources are given the classifications derived from the SED fits using the criteria described previously. For sources with {\it Herschel} 70 \micron\ fluxes, we found that the best fit SEDs consistently overestimated the 70 \micron\ fluxes, in most cases by more than an order of magnitude. Because long wavelengths are so important for constraining envelope properties, before we fit the SEDs of these sources, we upweighted the 70 \micron\ data to ensure that the best fits matched the observed 70 \micron\ fluxes. The resulting classifications are listed in Table \ref{tab:class}. Table \ref{tab:param} shows the results of the fitting for sources determined to be protostars. Here the columns list the source name, the value of $\chi^2$ for the best fit, number of data points in the SED ($N_{data}$),  the number fits that pass $\Delta\chi^2_{scaled}$ $<$ 1 criterion ($N_{fits}$), the number of fits classified as protostellar or disks ($N_{P}$ or $N_{D}$ respectively), and the median values of the ratio of derived envelope mass to stellar mass, the ratio of mass infall rate to stellar mass, and the stellar mass. 

\section{Analysis and Results}\label{results}

\subsection{Validation of SED Criteria\label{valid}}

\subsubsection{Synthetic YSO SEDs\label{synth}}

To determine the performance of the fitter and validate our criteria, we tested them with fake data generated from the \citet{robit2006} models with errors and wavelength coverage driven by the actual data. The error adopted in each band was the median error in that band for the Megeath sources. The synthetic sample consisted of 366 (175 Stage I and 191 Stage II) randomly selected models with parameters spanning nearly the full range of Robitaille's model grid, set at a distance of 390$\rm~kpc$. The input models were designated as either Stage I or II using the definitions described earlier. We applied a foreground extinction to each source. Extinctions were drawn from a uniform distribution in $\av$ from $0 - 10$ magnitudes. For each SED, we allowed each photometric point to vary by $3\sigma$ to simulate likely systematics which exist in the data (e.g., variability, calibration errors). Table \ref{tab:conf} presents the results of this experiment as a confusion matrix which shows the number of sources that were originally (row) Stage X and are classified as (column) Y. The row and column totals are shown on the side and bottom, respectively.  We find that we correctly classify 86\% of synthetic Stage I sources as protostars, and that 86\% of objects we classified as protostars were originally true (synthetic) Stage I objects. The fact that these values match is completely coincidental. Furthermore, the original ratio of Stage I to Stage II sources is preserved in the final classifications derived from our methodology.

The individual parameters (e.g., stellar mass, envelope mass, infall rate) derived from both the median and best fits to the synthetic data generally agree with the input values to within factors of 1.5 to 8 with the median values being better constrained.  This is in agreement with what was found by \cite{offner2012} but indicates that the individual source parameters derived from the \citet{robit2007} fitter must be regarded with some caution. However, we note that the derived ratios of these same physical parameters are somewhat more trustworthy, generally being within factors of 3-5 of the true values for both the best and median fits. The level of uncertainties in the individual physical parameters derived from the SED modeling may be of some concern. It is likely that the same parameters derived from the fits to the actual data are even more uncertain than suggested here. The magnitudes of $\chi^2$ for the fits to the real data are considerably greater than 1 and this already indicates that the standard models used here may not accurately represent the underlying physical structure of real YSOs.  Although the physical parameters derived from model fits are very uncertain, the results in Table \ref{tab:conf} indicate that the evolutionary classifications derived from the fitting appear to be reasonably reliable.

Among the 14\% of synthetic Stage I sources classified as Stage II or disk dominated objects, the fitted disks were predominantly highly inclined. This was particularly true for sources with lower mass envelopes. The presence of disks with
large intrinsic flaring in real observations could also affect derived classifications. Although the Robitaille models select flaring parameters such that disks and protostars would be distinct, some real YSOs might have larger flaring than assumed in the models. However, we expect that disk flaring will only affect the extinction for a short time, as the dust in the disk settles relatively fast compared to the long Class/Stage II lifetime  \citep{2006AJ....131.1574L}. Therefore flared disks should account for a only small fraction of real Stage II sources.

\subsubsection{Protostellar Candidates Associated with Outflows\label{outflows}}

Molecular outflows are a relatively short-lived phenomena typically associated with young protostellar systems and are indicators of youth. Outflows allow us to identify extremely young sources independent of their SEDs. Sources M1246, M2393, M2707, and M2748 of the Megeath low extinction protostar candidates have jets or outflows (see corresponding images in Appendix \ref{fig:images}). \citet{2009A&A...496..153D} used UKIDSS WFCAM to map $H_2$ at $2.122~\micron$ to trace molecular outflows and associated each outflow with a protostar from an earlier version of the Megeath catalog by tracing the jets back to their source. The outflows are labeled DFS\#. Sources M1246 and M2393 are associated with $H_2$ outflows DFS112 and DFS102 respectively.
In archival data from the Hubble Legacy Archive (HLA), a jet and terminating shock of at least $1\arcmin$ in length is visible for M2707 with Hubble's F656N filter. A bipolar jet associated with M2748 is visible in images from the VISTA Vienna Survey in Orion \citep[VISION\footnote{Available from: \url{http://homepage.univie.ac.at/stefan.meingast/vision.html}},][]{Meingast2015}. Each side of the outflow has a length of $\sim 20\arcsec$ (.05 pc projected physical length). Three out of these four sources were identified as protostars by our SED fitting analysis. The fourth, M1246 failed a protostellar classification due to too low an envelope mass derived from the SED fitting. Since, flat spectrum, disk dominated sources can drive jets, we retain the Stage II classification for this source. However, the criteria we used to define protostars captured 3 of these 4 outflow sources, providing some additional validation of our procedure. As we discussed earlier, the absolute values of the individual physical parameters returned by the SED fits are highly uncertain, in some cases uncertainties reach an order of magnitude.  Although, we note that if, as a definition for a protostar, we used the higher ratio of envelope to stellar mass of 0.1,  only two outflow sources would have passed the protostellar selection filter.

\subsubsection{Protostellar Candidates with Si Emission Features\label{sires}}

When available, 9-20 $\mu$m spectra spanning the solid silicate dust features are very useful for distinguishing between Stage I and II objects \citep{2008A&A...486..245C}. In particular, Si emission is known to arise in the inner regions of circumstellar disks where dust temperatures are high, whereas Si absorption arises in much colder circumstellar envelopes as well as in cold foreground, but unrelated, dusty material. Therefore, we assume here that the 9.7 $\mu$m Si emission arises from a circumstellar disk while absorption arises mostly from the protostellar envelope. This latter assumption is particularly appropriate for low extinction sources as those studied here. The presence of Si emission does not necessarily disqualify a source as a protostar. For example, when a protostar is observed with its outflow cavity along the line-of-sight, the circumstellar accretion disk and its Si emission feature are directly viewable. It may also be possible that in some circumstances the optical depth in the envelope is not sufficient to cancel and reverse the underlying emission feature from the disk. Nine protostellar candidates, \M333, \M632, \M845, \M892, \M931, \M1005, \M1474, \M2255 and \M2561 were found to exhibit Si emission in their IRS spectra. Eight of these (\M845, \M632 \M892, \M931, \M1005, \M1474,\M2255, \M2561) were classified as Stage II sources as a result of our SED analysis and the presence of Si emission confirms our identifications. The Si emission in the SED of M333 is effectively modeled with a face-on Stage I protostar. Since the model fits for this source are also consistent with the IRS spectra, our classification of M333 as a protostar is retained.

\subsubsection{Protostellar Candidates with Optical Emission Lines\label{optline}}

Four of the low extinction protostellar candidates in the Megeath catalog were spectroscopically classified as emission-line, PMS stars (i.e., T Tauri stars or CTTS) by \citet{fang2013}. Three of these sources -- M632, M845, and M892 are classified as Stage II sources by the SED analysis of this paper, consistent with the optical classifications. One other, M333, was classified as a protostar by our SED analysis and as a flat spectrum source by \citet{fang2013}, essentially consistent with our results.
 It is important to note that at short (i.e., optical and near-infrared) wavelengths CTTSs are not always distinguishable from young face-on protostars. Protostars are essentially actively accreting CTTS surrounded by an infalling envelope and if they have an outflow cavity directed along our line-of-sight, direct near-infrared and optical spectroscopy of the central, CTTS-like, photosphere is possible. Indeed, as mentioned above, we found the SED of source M333 to be best fit by a face-on protostellar model enabling both infrared Si and optical emission lines to be detected. The overall agreement of the optical spectroscopy with our assigned classes provides additional evidence in support of our method of classification.

\subsubsection{Foreground and extragalactic contaminants\label{exgal}}

Since star-forming galaxies and AGN are isotropically distributed and have colors matching YSOs, they are an ever present source of contamination. \citet{megeath2012} predict $1.5\pm1.5$ contaminating objects/deg$^2$ misidentified as protostars. Orion has $6$ square degrees ($\sim320\rm\ pc^2$) within the Megeath survey region below 0.8 mag $\ak$, which means we would expect $9\pm9$ extragalactic or field/foreground sources contaminating our list. We find four galaxy contaminants, \M623, \M742, \M1283, and \M2255. To identify galaxies, we examined their SEDs, images from the VISION survey, and their IRS spectra where available. The image of source \M623 shows two overlapping extended red objects, one of which is an elongated ellipsoid. If this were a disk, it would be seen as a shadow against scattered light instead of in emission as it is. In addition, the absorption feature we note at 9.7$\mu m$ is very narrow and imposed on a straight, rising spectrum that shows PAH emission and is similar to other galaxy spectra (e.g. \cite{brandl2006}). The spectra of \M1283 and \M623 are nearly identical. \M2255 shows emission at 9.7$\micron$ and is also very flat and rising. The PAH and other galaxy features are weaker than in \M623 and \M1283, but are still apparent. \M742 has an SED which is not rising steeply enough for it to classified as Class I source and it is classified as Stage II by the fitting criteria but it is not particularly well fit by the \citet{robit2006} models. In the VISTA images, it is extended and asymmetric. If this were a face-on disk at the distance to Orion, it would be about 800 AU in extent (in scattered light) and its SED would appear more stellar-like than it does at short wavelengths. For these reasons, it is classified as a galaxy. \M1005 is identified as a galaxy by the HOPs team \cite{furlan2016}, however  \cite{furlan2016} note that it could be a so-called transition disk, which is consistent with our classification as a Stage II YSO.

\subsection{The Low Extinction Protostellar Population \label{lepp}}

As summarized in Tables \ref{tab:class} and \ref{tab:stat}, we find that out of \samplesize\ candidate low extinction protostars \stageI\ (\pstageI\%) are classified as protostars by our more restrictive classification criteria, while \stageII\ (\pstageII\%) are Stage II/Class II sources, including three of four previously classified as CTTS. Additionally we find that \fuzz\ (\pfuz\%) are extended nebulous regions and cloud edges misidentified as stellar-like sources. Four (\pgal\%) of the sources we suspect of being extragalactic contaminants. Finally, there are \unk\ (\punk\%) sources which remain unclassified. We ran the same analysis on all 285 high extinction ($\ak>0.8$ mag) candidate protostars in the Megeath catalog as a test of our criteria. Only 4\% (11) of the 285 protostellar candidates at high extinction had too little data to enable classification by our criteria. By selecting $10^5$ random draws of 44 sources from the sample of 274 we could classify, we find that for a given draw we classify $50\pm7\%$ of the sources as protostars, and the rest as Stage II, by our stringent criteria.  Our result is consistent with other studies that use methods other than color and spectral indices to classify YSOs in other star forming regions. For example, using a classification scheme similar to ours, \citet{forbrich2010} reclassified $20\%$ of protostellar sources in NGC 2264 from Class 0/I to Stage II. \citet{2009A&A...498..167V} studied Class I objects in Ophiuchus with molecular line and submillimeter continuum observations and reclassified 23/45 (51\%) of the sources with Class I SEDs as edge-on disks.  Of the 41 protostellar candidate sources in low extinction regions that we were able to classify 10\ (24\%) appear to be likely true protostars. This perhaps could be considered as a lower-limit to the actual population at low extinction since we have not included possible protostellar objects that may have been misclassified as Class II objects by \citeauthor{megeath2012} However, we expect the numbers of such objects to be small since, as can be inferred from  Figure \ref{fig:med_all}(b), the Megeath et al. criteria for Class II/III sources appears to very effectively screen out Stage I objects with protostellar (Class I) SEDs. However, the converse, that the Megeath et al.  criteria for Class I sources effectively screen out Stages II objects, is clearly not true. 

In terms of spatial distribution, most (90\%) of the confirmed low extinction protostars are located around the integral shaped filament in the Orion Nebula. This is higher than the fraction (70\%) of all the \citeauthor{megeath2012} candidate low extinction protostars also located in the same region. Figure \ref{fig:small} shows this region and the locations of the confirmed and candidate low extinction protostars within it. Only one confirmed protostar is found in the vast area of low extinction within the cloud but outside this region (see Fig. \ref{fig:all}).

\begin{figure}[htbp]
\plotone{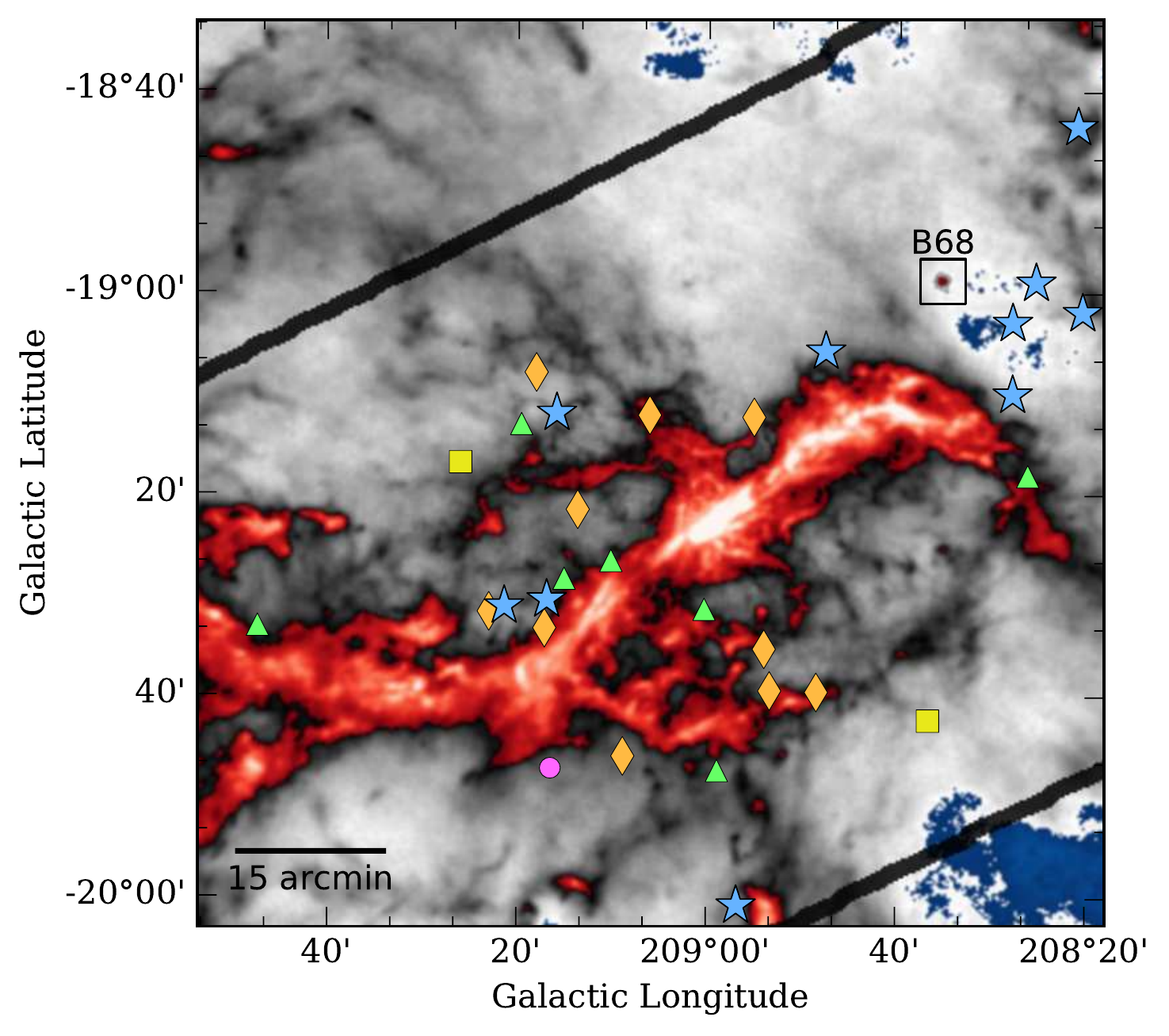}
\caption{\label{fig:small}{\it Herschel}\footnotesize\ K-band extinction ($\ak$) map of the Integral Shaped Filament (ISF) in Orion A with low extinction sources overlaid. The insert shows the extinction map of the Bok globule B68 overlaid on the Herschel map. It is clear that a dense core with a similar mass ($\sim$ 1-2 $\msun$) and size would easily be detected in the low extinction (grey) regions of the Herschel map. The thick black lines represent the boundaries of the \protect\citet{megeath2012} survey region.  Otherwise same as Figure 2. }
\end{figure}

\subsection{Protostellar Impostors but Still Quite Young\label{impostors}}

 In Figure \ref{fig:med_low}, we separated the \citeauthor{megeath2012} candidate low extinction protostars into confirmed protostars and disk dominated YSOs, as we classify them. The shape of the median protostellar SED matches both general expectations for Class 0/I sources as well as the shape of the median SED for all Megeath protostars in Orion A (Figure \ref{fig:med_all}). However, the candidate protostars we reclassified as Stage II sources still exhibit SED shapes more similar to Class I protostellar objects than to Class II objects (e.g., Figure \ref{fig:med_all}). This likely indicates that this group of Stage II objects represents an extremely young population of disk dominated YSOs which still are associated with material from either remnant circumstellar envelopes or highly flared disks. Indeed, most of these sources are flat spectrum sources which are often considered to be sources in the very late stages of protostellar evolution in transition to the Class II phase \citep[e.g.,][]{greene1994}. From examination \citeauthor{megeath2012}'s measurements of the 2-8 micron SED slopes ($\alpha_{IRAC}$) of low extinction sources, we find only 3\% (29 of 962 sources with measured slopes) of their class II objects to be flat spectrum sources (i.e., -0.3 $\le$ $\alpha_{IRAC}$ $\le$ $+$0.3). Including the objects we reclassified to Stage II, we estimate that less than about 5\% of the Stage II objects at low extinction in the Orion A cloud are in this flat-spectrum category, confirming the notion that these sources, though not true protostars, are likely still extremely young. If the typical Stage II lifetime is $\sim$ 2 Myr, then the flat spectrum phase of Stage II evolution has a duration of only about 10$^5$ yrs., emphasizing the extreme youth of such objects. 

\begin{figure*}[htb]
\centering
\includegraphics[width=5in]{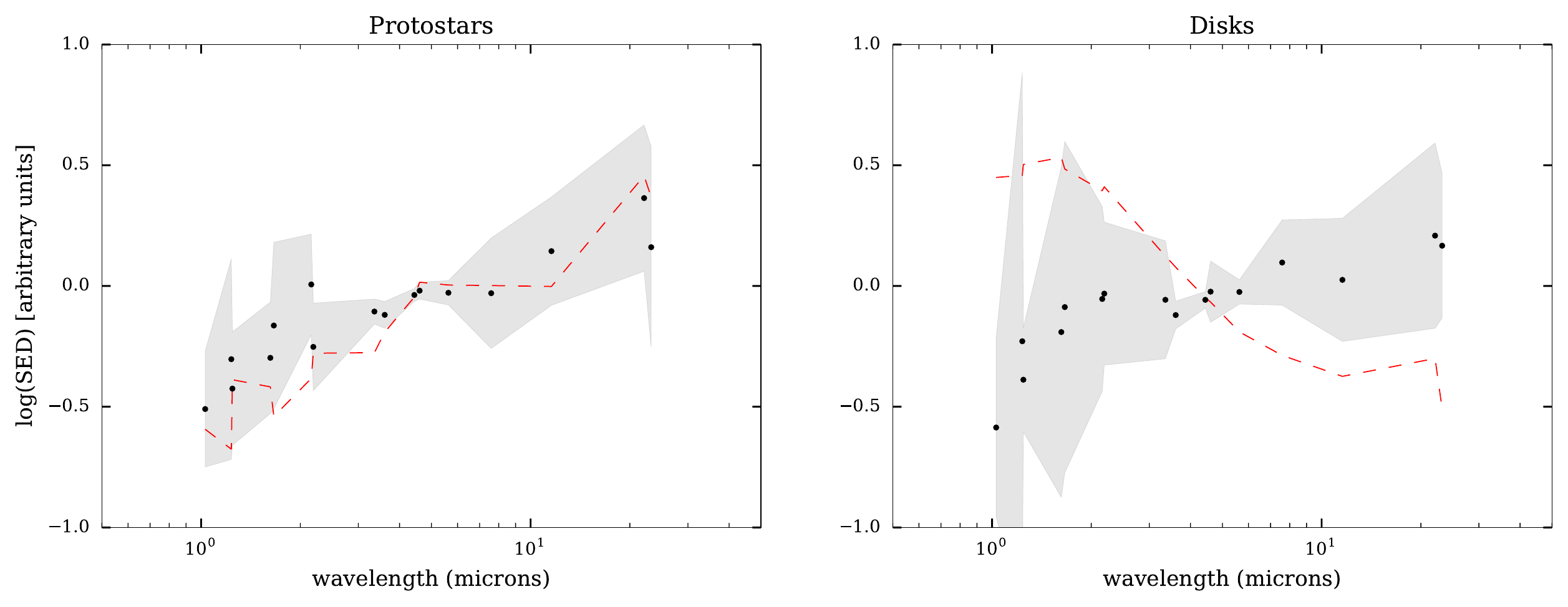}
\caption{\footnotesize\label{fig:med_low} Median spectral energy distribution of sources we have classified as protostars (left) and disk-dominated Stage II objects (right) at low extinction. The red trace shows the corresponding median SEDs from Figure \ref{fig:med_all}. The spectral slopes from 2-8 microns are $0.21\pm0.2$ and $0.19\pm0.13$ for the protostar and disk median SEDs respectively. The median SED of reclassified protostars (right) differs significantly from the the median SED of Class II sources in Orion suggesting the reclassified sources are Stage II sources which are likely extremely young (see text, \S\ref{impostors}).} 
\end{figure*}

 Finally, we note the intriguing result that the slope we determine for the median SED of the entire Megeath et al. Class II population in Orion A (Figure \ref{fig:med_all}b) has a value of -1.33. This is surprising as it is the same as that expected for a spatially flat, classical accretion (or reprocessing) disk, although the nearly exact agreement is likely coincidental. The relatively steep slope we measure suggests a high degree of dust settling has occurred in a significant fraction of the disk sources in the Orion A cloud. This finding is similar to that obtained in an earlier study of the disk population of the IC 348 cluster \citep{2006AJ....131.1574L} and emphasizes the already well known fact that disk evolutionary timescales are relatively short compared to the PMS timescales of low mass stars \citep[e.g.,][]{2001ApJ...553L.153H}.

\section{Discussion}\label{discuss}

With \stageI\ confirmed protostars in low column density gas, we now consider the interesting question of why they are found at low extinction. Is it possible for stars to form in situ in such an environment or are there other reasons for their presence in low extinction regions? We discuss here possible scenarios to explain their presence in low extinction gas.

\subsection{In Situ Formation\label{insitu}}

Observational and theoretical considerations have long suggested that low mass (i.e., 1-5 $\msun$) dense cores are the sites of formation of individual low mass stars \citep[e.g.,][]{1986ApJ...307..337B, 1987ARA&A..25...23S,1990ApJ...365L..73Y} while more massive (20-2000 $\msun$) dense cores are the formation sites of stellar groups and clusters \citep{1992ApJ...393L..25L}. If the low extinction protostars we have identified in Orion were formed close to their present positions, we might expect them to be physically associated with small, relatively isolated, dense cores within an otherwise more extended, low extinction portion of the cloud. Given the distance of Orion and the resolution of our Herschel maps, could we detect such a dense low mass core in the Herschel data? To address this question consider the well studied starless dark core, B68. It probably represents the lowest mass dense core that would surround an isolated protostar. It is a well defined and isolated Bok globule with a mass of $2.1\ \msun$ and a $12500\rm\ AU$ radius \citep{2001Natur.409..159A}. At the distance of Orion (420$\pc$), B68 would have an angular diameter of $\sim50\arcsec$, so it would be resolved at the {\it Herschel} extinction map resolution of $36\arcsec$. To illustrate how B68 would appear when located within the Orion molecular cloud, we convolved the infrared extinction map of B68 from the survey of \citet{2010ApJ...725.2232R}
with the {\it Herschel} beam, regridded it to the Herschel extinction map of Orion and placed it in Figure \ref{fig:small}. As can be seen, B68 would be readily detectable, displaying a peak extinction of $\ak=1.5$ mags. At $R=30\arcsec$ the extinction drops to $\sim0.5\ak$. Clearly none of our protostars are currently within cores such as B68. Because a core which recently formed a protostar would be characterized by even higher peak extinctions than B68, we must conclude that the low extinction protostellar objects are orphan protostars that have lost their original cores either because they were formed elsewhere in high extinction regions and migrated to their present positions or because through the course of their evolution have had their surrounding cores removed or dissipated in place.

\subsection{Source Migration/Ejection\label{ejection}}

Direct measurements of core and protostellar kinematics have suggested that embedded protostars remain closely associated with dense cores during their lifetimes. \citet{2006AJ....131..512C} derived an upper limit ($\lesssim 1.8\kms$) to the protostellar velocity dispersion from spectroscopic observations of protostars in several nearby clouds. This upper limit was comparable to the observed velocity dispersion of CO in these clouds indicating that the protostars were still dynamically coupled to at the least the bulk gas in the clouds. However, this dispersion is higher than the typical trans-sonic velocity dispersions ($\sim 0.33 - 0.55 \kms$) measured for dense cores in local clouds (e.g.,  \citet{2007ApJ...671.1820M};   \citet{2010ApJ...723..457K}). Although the difference between the upper bounds on the protostellar dispersion and the measured core velocity dispersions could be consistent with the presence protostellar migration, it is unclear how significant this difference really is. Because the measured protostellar velocity dispersion can be mostly accounted for by systematic uncertainties, the true value is likely much smaller \citep{2006AJ....131..512C}.  Therefore it seems reasonable to conclude that the velocity dispersion of protostars is similar to that of their host cores and that during their lifetimes most protostars do not drift much beyond their parental cores. A limit on protostellar motions in Orion A has been estimated by \cite{2014A&A...566A..45L} using a Bayesian technique. \citeauthor{2014A&A...566A..45L} model the relation between the surface densities of protostars and dust and simultaneously fit for a number of parameters including a spatial diffusion coefficient ($\sigma$) for protostars using the Megeath catalog and {\it Herschel}-derived extinction maps. They determine $\sigma=0.0088\pm 0.0049\pc$. Such a small $\sigma$ suggests that on average there is not significant migration of the protostellar population in Orion. This is perhaps consistent with our finding that roughly 90\% of the protostars in Orion are presently within the highest extinction regions of the cloud.

Nonetheless it is possible that the during their lifetimes the low extinction protostars in Orion have moved from their birth sites in the high extinction regions of the cloud. To help assess the feasibility of this scenario, we estimate the velocities required for each protostar to reach the nearest high extinction region within an average protostellar lifetime of $2\times10^5$yr. We determine what velocity is required for our objects to migrate to their current locations from the $0.5$ and $0.8$ $\ak$ extinction contours. The projected physical distances from the nearest $0.8\ak$ and $0.5\ak$ contours, and corresponding velocity estimates are listed in Table \ref{tab:cont}. The predicted velocities range from $0.72-8.1\kms$ for $v_8$ and $0.01 - 0.88\kms$ for $v_5$.

As noted earlier, (see Figure \ref{fig:small}) the bulk of the low extinction protostars are located near the so-called integral filament. This is also the region where most of the stars in the cloud, including the Trapezium cluster, have formed. Simulations have suggested that dynamical interactions between stars in a forming cluster can result in ejection of a small fraction of the cluster stars \citep{2003MNRAS.339..577B}. The simulations show that ejected stars can reach velocities up to $\sim1 \kms$ and travel
distances of 0.2$\pc$ in two hundred thousand years \citep{2003MNRAS.339..577B}. This is much smaller than the projected distances of the low extinction protostars from the center of the Trapezium cluster (2 - 20 pc). To reach the observed locations of the low extinction protostars from the Trapezium cluster, ejection velocities between 10-50 $\kms$ would be required. Ejection from the cluster is therefore a highly unlikely cause of the presence of these protostars at low extinction.
Although migration due to dynamical ejection from the cluster seems highly improbable, only direct measurements of the protostellar velocities will ultimately determine whether such a process has occurred in Orion.

Recently, \cite{stutz2016} have argued that once a protostar has accreted a substantial fraction of its final mass it can kinematically decouple from the dense material in which it formed. They posit that bulk motions of the dense gaseous filaments (which contain the protostellar cores) can leave the decoupled protostars behind with relative velocities as high as 2.5 $\kms$. \cite{stutz2016} further posit that such large ejection velocities might explain the larger velocity dispersion of Class II sources compared to that of protostellar cores in Orion. However, recent observations described by  \cite{hacar2016} suggest such velocity dispersions of Class II sources may inherently overestimate any real protostar-core velocity drift because these dispersions may represent averages over multiple, individually coherent velocity structures (e.g., strings or filaments) along the line-of-sight. Nonetheless, such ejection of protostars as suggested by \cite{stutz2016} would ultimately and rapidly quench accretion of material onto the protostellar surface since the protostar would be removed from the gas supply of its parental core. In the earliest stages of such an ejection process the protostars might resemble those we observe in the low extinction regions, that is, coreless protostars with low accretion luminosities and envelope to stellar mass fractions. However, whether this is truly a viable mechanism to explain the existence of the low extinction protostars in Orion remains to be seen. Measurements of proper motions with Gaia along with direct radial velocity measurements of these particular sources will provide more stringent tests of this idea.

\subsection{Excavation: Evolution In Place\label{excavation}}

Protostellar outflows have the ability to significantly affect the local environment in a giant molecular cloud \citep{2014prpl.conf..451F}. The outflow creates an outflow-blown cavity which expands in the cloud. The smooth expansion is disrupted by the turbulence remaining in the cloud, transferring energy from the outflow to the cloud \citep{2007ApJ...659.1394M, 2009ApJ...692..816C, 2005ApJ...632..941Q, 2010ApJ...715.1170A}. While the outflow is often not energetic enough to unbind the cloud, it does inject enough turbulence into the cloud to provide some support against collapse \citep{2009ApJ...695.1376C, 2007ApJ...659.1394M}. In NGC 1333 the contribution of outflows alone is believed to be enough to maintain the turbulence and prevent further collapse \citep{2013ApJ...774...22P, 2007ApJ...659.1394M}. If this process occurs at the edge of a cloud, where the density in steadily decreasing, it's not hard to imagine that bipolar jets could carve out a cavity and disperse the gas since it is less bound than in the densest portions of the cloud where the \citet{2013ApJ...774...22P} and \citet{2007ApJ...659.1394M} studies apply.

The environments around M1246 and M2393 are suggestive of possible cloud excavation by jets. Both of these sources have outflows about .3 pc in length identified by \citet{2009A&A...496..153D}. The dust morphology around \M2393 morphology is suggestive of a cavity (Fig.~\ref{fig:2393}). The cavity lies along the axis of the outflow on the opposite side of protostar. The jet from \M1246 goes through a gap in the filamentary structure to its galactic north east (Fig.~\ref{fig:1246}). M2748 is located in a large structure reminiscent of a core. If this were a core that was disrupted, given the surface density of the material, only material within 0.1 pc with $\ak>0.35$,
in the bowtie shaped structure (Fig.~\ref{fig:2748}) around the protostar, would be required to make a core as massive as B68. Two other low extinction protostars, M2707 and M2393 have outflows but no obvious evidence for cavities or interaction with surrounding material in the Herschel images, although M2707 is close to the edge of a dense core-like structure. The remaining five low extinction protostars apparently do not have outflows at present. Nonetheless the sources could be sufficiently evolved that outflow activity has declined to the point that any jets present are not so easily detectable, or they may undergo episodic accretion \citep[e.g.,][]{2010ApJ...710..470D} and be in a quiescent state.

In an interesting recent study of the YSO population in the main core of the Rosette Molecular Cloud (RMC), \citet{2013ApJ...769..140Y} found that the level of extinction toward a YSO appears to be systematically correlated with its estimated age.  Specifically, \citet{2013ApJ...769..140Y} showed that the extinction, $\ak$, toward the YSOs in the RMC cloud decreased at an average rate of $0.5~\rm mag ~Myr^{-1}$. \citeauthor{2013ApJ...769..140Y} suggest that this is the combined effect of the initial star formation event being spread throughout the nebula and of the rapid removal of gas from around the individual objects. On the Class I timescale ($2\times10^5$yr), $\ak$ would be expected to change by 0.1 mag. If such a process were at work in Orion this would place sources M0333, M0892, M2561, and M2748 close to $\ak\sim0.8$ mag at the time of their formation, assuming that they were all in the late stages of protostellar evolution. This evolution could be driven by either migration or local excavation of surrounding gas.

\section{Conclusion}\label{conclus}

We find that 24\% of the candidate low extinction ($\ak<0.8$\ mags) protostars from \citet{megeath2012} are confirmed as protostars according to our more stringent criteria. At high extinction, where we might expect the existing classifications to be better, we find $>50\%$ of the \citeauthor{megeath2012} Class 0/I sources pass our protostellar criteria. The rate is much lower at low extinction, suggesting that there is a real increase in the fraction of incorrectly identified protostars at low extinction in the \citeauthor{megeath2012} catalog. We find that while simple color selection criteria may be adequate for describing statistically large samples of YSOs, they are not always able to distinguish Stage I (protostellar) from Stage II (disk) sources with flat or Class I SEDs. More accurate determination of the evolutionary stage of a source requires SED modeling with long wavelength data that can better probe the infalling envelope. 

Although SED modeling can be an effective tool for classifying protostellar sources, we find that the individual physical parameters (envelope mass, accretion rate, stellar mass, etc.) determined by SED-fitting can be very uncertain and should be treated with appropriate caution. Derived ratios between certain parameters may be somewhat more reliable.  Near-infrared spectroscopy \citep{lw1984,greene1996,furlan2008,forbrich2010} and submillimeter, molecular-line observations \citep{2009A&A...498..167V,2008A&A...486..245C} would be extremely useful for a more confident verification of a source's status as a protostellar object.

We find that none of the low extinction protostars in the Orion A cloud are presently embedded in solar mass dense cores that are typically associated with low mass star formation. However, the confirmed low extinction protostars do not necessarily have to be formed in situ from low extinction material, although this possibility cannot be firmly excluded by existing data. Instead, we posit that these protostars are orphans that have lost their original parental cores either via migration/ejection from higher density regions or through dissipation due to evolutionary processes such as outflow activity and accretion of original material onto the star. Given the short protostellar lifetimes, it seems highly unlikely that migration or ejection from existing high extinction regions can account for the low extinction protostars, at least for the majority of cases. Excavation due to the actions of outflows may be the most viable option for explaining the presence of these objects in extended regions of low extinction gas and dust. The lack of direct evidence for outflow activity in 5 of the 9 confirmed protostars is somewhat troublesome but may indicate that these objects either experience episodic accretion, and are in a quiescent state, or are in a late stage of protostellar evolution. 

Additionally we note that the presence of orphan protostars projected on low extinction gas does not necessarily preclude the existence of an extinction threshold for star formation. The precise location of that threshold is still somewhat uncertain but probably lies above 0.5-0.8 magnitudes of K-band extinction. Even with our more stringent classification criteria we estimate that 93\% and 95\% of all protostars are located in regions with A$_K$ $>$ 0.8 and 0.5 magnitudes, respectively. Furthermore from the fact that the gaseous masses in the Orion cloud are roughly 1.32 $\times$ 10$^4$ and 5.44 $\times$ 10$^4$ $\msun$ respectively above and below the extinction threshold of 0.8 magnitudes, we find the corresponding efficiencies of protostar formation to be 1.8 $\times$ 10$^{-4}$ and 1.0 $\times$ 10$^{-2}$ protostars $\msun^{-1}$, respectively. That is, in Orion A the yield of protostars per unit cloud mass is approximately 58 times higher in regions above 0.8 magnitudes of extinction than in regions below. 

Finally, we note that although our more restrictive classification criteria found only 36\% (10/28) of the  viable \citeauthor{megeath2012} protostellar candidates to be protostars, the remaining objects, though now designated as Stage II objects, still do not have colors or SED signatures that are typical of most known Class II sources. Our newly classified Stage II sources have mostly flat-spectrum SEDs and are likely in the very earliest phases of Stage II/PMS evolution and therefore are still relatively young objects.

\acknowledgements
{\it\noindent Acknowledgments\\}
{\footnotesize
We thank Stefan Meingast and Jo\~ao Alves for the VISTA VISION survey images, Marco Lombardi for the {\it Herschel} data and Josefa Gro\ss schedl for very useful discussions. We thank the anonymous referee for comments and criticisms that greatly improved the quality of this paper. JAL acknowledges useful discussions with Phil Cowperthwaite, Dawn Graninger, Alyssa Goodman, John Johnson, Karin \"Oberg, and Ryan Loomis. JAL was funded by the Harvard Graduate Student Fellowship, the Dick Smith Family Fellowship administered by Harvard University, and a NSF Graduate Student Research Fellowship. This research has made use of NASA's Astrophysics Data System; the NASA/IPAC Infrared Science Archive; the VizieR catalogue access tool, CDS, Strasbourg, France; Astropy, a community-developed core Python package for Astronomy \citep{2013A&A...558A..33A}; and APLpy, an open-source plotting package for Python hosted at http://aplpy.github.com.

\clearpage
%%%%%%%%%%%%%%%%%%%%%%%%%%%%%%
%%%%%% TABLES %%%%%%%%%%%%%%%%
\begin{deluxetable}{llcccccc}
\tabletypesize{\scriptsize}
\tablewidth{0pt}
\tablecaption{Low $\ak$ Source Classification \label{tab:class}}
\tablehead{
 \colhead{Source ID} &
 \colhead{$\ak$(err)} &
 \colhead{Si feature} &
 \colhead{Classification} &
 \colhead{Notes} & 
 \colhead{HOPS ID$^\text{(4)}$} & 
 \colhead{HOPS Class$^\text{(4)}$} &
 \colhead{Ref}}
\startdata
0125 & \akv{0.53}{0.02} & & Stage II & & & \\
0200 & \akv{0.56}{0.02} & & Stage II & & & \\
0333 & \akv{0.71}{0.02} & Emission & Protostar & CTTS & HOP-230 & Flat  & 2 \\
0564 & \akv{0.12}{0.01} & & Stage II & & \\
0623 & \akv{0.73}{0.04} & Absorption & Galaxy & & HOP-161 & Galaxy &  \\
0632 & \akv{0.23}{0.01} & Emission & Stage II & CTTS, Transition Disk & HOP-126 & Flat & 2 \\
0730 & \akv{0.08}{0.01} & & Stage II & & \\
0740 & \akv{0.40}{0.01} & Absorption & Stage II & WTTS & HOP-147 & Flat & 2b \\
0742 & \akv{0.10}{0.02} & & Galaxy & &\\
0845 & \akv{0.72}{0.03} & Emission & Stage II & CTTS ; Binary & HOP-170 & Flat & 3 ; HST \\
0892 & \akv{0.66}{0.02} & Emission & Stage II & CTTS & HOP-199 & Flat & 2 \\
0931 & \akv{0.56}{0.02} & Emission & Stage II & WTTS, Thick Disk & HOP-187 & Flat &  2,4 \\
1005 & \akv{0.10}{0.01} & Emission & Stage II & &  HOP-202 & Galaxy &  \\
1042 & \akv{0.51}{0.02} & Absorption & Stage II & & HOP-195 & Flat  & \\
1246 & \akv{0.26}{0.01} & & Stage II & Jet source - DFS 112 &  HOP-25 & Flat &  1\\
1283 & \akv{0.27}{0.01} & Absorption & Galaxy &  & HOP-27 & Galaxy &\\
1316 & \akv{0.38}{0.02} & & Protostar &  & HOP-31 & Flat &\\
1355 & \akv{0.56}{0.02} & & Stage II &  & HOP-382 & Class I & \\
1369 & \akv{0.32}{0.01} & & Protostar & & & \\
1431 & \akv{0.32}{0.01} & & Stage II & &  & \\
1472 & \akv{0.70}{0.06} & & Stage II & &  & \\
1474 & \akv{0.42}{0.02} & Emission & Stage II & & HOP-47 & Flat  &\\
1512 & \akv{0.32}{0.02} & & Protostar &  & & \\
1662 & \akv{0.59}{0.02} & & Stage II &  & & \\
2255 & \akv{0.18}{0.01} & Emission & Galaxy & & HOP-61 & Galaxy & \\
2393 & \akv{0.27}{0.01} & Absorption & Protostar & Jet source - DFS 102 & HOP-83 & Galaxy &  1 \\
2561 & \akv{0.59}{0.01} & Emission & Stage II & & HOP-102 & Class I & \\
2593 & \akv{0.16}{0.01} & & Protostar & & HOP-104 & Class I & \\
2649 & \akv{0.12}{0.01} & No Features & Protostar & & HOP-105 & Flat & \\
2690 & \akv{0.13}{0.01} & & Protostar & &  & \\
2707 & \akv{0.26}{0.03} & No Feature & Protostar & Jet source & HOP-107 & Flat & \\
2748 & \akv{0.65}{0.02} & & Protostar & Jet source  & & \\
\cutinhead{Other (12)}
0122 & \akv{0.52}{0.01} & & Unknown & point source & \\
1203 & \akv{0.39}{0.01} & & Unknown & point source & \\
1220 & \akv{0.23}{0.01} & & fuzz & & \\
1286 & \akv{0.72}{0.03} & & fuzz & & HOP-381 & Uncertain & \\
1313 & \akv{0.38}{0.01} & & fuzz & & \\
1453 & \akv{0.28}{0.01} & & fuzz & & \\
1513 & \akv{0.20}{0.01} & & fuzz & & \\
1760 & \akv{0.47}{0.02} & & cloud edge & & \\
1839 & \akv{0.49}{0.02} & & cloud edge & & \\
1858 & \akv{0.61}{0.03} & & fuzz & & \\
1983 & \akv{0.68}{0.03} & & fuzz & & \\
2221 & \akv{0.37}{0.04} & & Unknown & poorly fit & &
\enddata
\tablerefs{(1) \citet{2009A&A...496..153D}; (2) \cite{fang2013}; (3) \citet{fang2009}; (4) \citet{furlan2016}; (5) \citet{2012ApJ...752...59H}}
\tablecomments{The criteria passed by each source are listed. We classify the Si[9.7$\mu m$] feature as being in Emission, Absorption, or No Feature if there is no clear sign of emission or absorption. We note previous classifications of these sources and which ones have jets. "cloud edge" sources are associated with more structure than "fuzz" which are essentially blank fields in the NIR. }
\end{deluxetable}

\begin{deluxetable}{ccccccccc}
\tablewidth{0pt}
\tablecolumns{9}
\tablecaption{Derived Parameters for Protostars\label{tab:param}}
\tablehead{\colhead{Source} &
    \colhead{$\chi^2_{best}$\tablenotemark{\it (a) }} &
    \colhead{~~$N_{data}$\tablenotemark{\it (b)}} &
    \colhead{~~$N_{fits}$\tablenotemark{\it (c)}} &
    \colhead{~~$N_{P}$\tablenotemark{\it (d)}} &
    \colhead{~~$N_{D}$\tablenotemark{\it (e)}} &
    \colhead{~~${M_{env}}/{M_{\star}} $} &
    \colhead{${\dot{M}}/{M_{\star}}  $}  &
    \colhead{$M_{\star}$} \\
    \colhead{} &
    \colhead{} &
    \colhead{} &
    \colhead{} &
    \colhead{} &
    \colhead{} &
    \colhead{ ($\times 10^{-2}$)} &
    \colhead{($\times 10^{-6}\ \text{yr}^{-1}$)} &
   \colhead{($M_\odot$)}
}
\startdata
333 & 145.8 & 7 & 7 & 5 & 2 & 15.5 & 12.0  & 0.2   \\
1316 & 382.4 & 12 & 10 & 6 & 4 & 17.1  & 28.4  & 0.2   \\
1369 & 8.4 & 5 & 4 & 3 & 1 & 83.5  & 92.9 & 0.3   \\
1512 & 67.9 & 8 & 17 & 17 & 0 & 59.2  & 80.2 & 0.2   \\
2393 & 15587.7 & 19 & 35 & 23 & 12 & 69.6 & 67.7  & 0.4   \\
2593 & 718.0 & 11 & 36 & 30 & 6 & 131.0  & 162.5  & 0.2  \\
2649\tablenotemark{\dagger}  & 4883.2 & 16 & 2 & 1 & 1 & 7.2 & 9.4 & 1.4 \\
2690 & 2934.4 & 13 & 7 & 5 & 2 & 17.3  & 30.5  & 0.2  \\
2707\tablenotemark{\dagger} & 10446.4 & 17 & 4 & 3 & 1 & 7.9 & 5.2 & 0.5 \\
2748 & 8388.1 & 20 & 21 & 18 & 3 & 27.5  & 17.5  & 0.2 
\enddata
\tablecomments{Weighted median model parameters for models with $\Delta\chi^2_{scaled} \le 1$. The uncertainties on the parameters are as discussed in \S\ref{results}. $^\dagger$ Sources \M2649 and \M2707 fit parameters were derived with the estimated {\it Herschel} 70 $\micron$ photometry from \protect\citet{furlan2016}.\\  {$^{(a)}$} $\chi^2_{best}$ is the $\chi^2$ of the best fit model (i.e. the lowest $\chi^2$ value). {$^{(b)}$}$N_{data}$ is the number of data points used for the SED fit. {$^{(c)}$}$N_{fits}$ is the number of fits which pass the $\chi^2_{scaled}$ criteria. {$^{(d,\,e)}$} $N_P$ and $N_{D}$ are the number of Stage I or II (respectively) models which pass the $\chi^2_{scaled}$ criteria. }
\end{deluxetable}

\begin{deluxetable}{l|cc|c}
\tablecaption{Classifier Confusion Matrix: Low Extinction \label{tab:conf}}
\tabletypesize{\footnotesize}
\tablewidth{0pt}
\tablehead{ & Protostar & Disk & Total}
\startdata
Stage I  & 150 & 25  & 175 \\
Stage II & 24  & 167 & 191 \\ \hline
Total    & 174 & 192 & 366 \\
\enddata
\tablecomments{The data show the number that correspond to a particular (row,column) pair. The terminal row and column show the totals for their respective row and column The rows are the actual source stage. The columns are the source stage determined using the $\chi^2$ weighted median of the fit parameters with $\Delta\chi_{scaled}^2 \le 1$, our classification scheme.}
\end{deluxetable}

\begin{deluxetable}{cccc}
\tabletypesize{\footnotesize}
\tablewidth{0pt}
\tablecaption{Classification Statistics \label{tab:stat}}
\tablehead{
\colhead{Type} & \colhead{Number}}
\startdata
Protostar & 10 (24\%) \\
Stage II  & 18 (44\%)\\
Galaxy    & 4  (10\%)\\
Fuzz      & 9 (22\%)\\ \hline
Total     & 41 (100\%) \\ \hline
Unknown    &  3
\enddata
\end{deluxetable}

\begin{deluxetable}{cccccc}
\tabletypesize{\footnotesize}
\tablewidth{0pt}
\tablecaption{Distance/velocity from .8 and .5 $\ak$ contours}\label{tab:cont}
\tablehead{
	\colhead{Source}
	& \multicolumn{2}{c}{$.8\ak$ contour} &
	& \multicolumn{2}{c}{$.5\ak$ contour} \\ \cline{2-3} \cline{5-6}
	 \colhead{ }
	& \colhead{$\rm d_8[\pc]$}
	& \colhead{$v_8[\kms]$} 
	& \colhead{}
	& \colhead{$\rm d_5[\pc]$\tablenotemark{\dagger}}
	& \colhead{$v_5[\kms]$}
	}
\startdata
 333 & 0.09 & 0.44 & & \multicolumn{2}{c}{\nodata} \\ %-0.17 & 0.83 \\
 1316 & 0.15 & 0.73 & & 0.03 & 0.15 \\
 1369 & 0.16 & 0.78 & & 0.11 & 0.54 \\
 1512 & 0.47 & 2.3 & & 0.1 & 0.49 \\
 2393 & 0.33 & 1.61 & & 0.18 & 0.88 \\
 2593 & 0.4 & 1.96 & & 0.31 & 1.52 \\
 2649 & 1.05 & 5.13 & & 0.89 & 4.35 \\
 2690 & 1.61 & 7.87 & & 0.6 & 2.93 \\
 2707 & 1.69 & 8.26 & & 0.04 & 0.2 \\
 2748 & 1.1 & 5.38 & & \multicolumn{2}{c}{\nodata}  %-0.04 & 0.2 \\
\enddata
\tablecomments{2D distances and velocities to the 0.5 and 0.8 $\ak$ contours. Velocities are estimated as $v=\sfrac{d}{(2\times10^5 {\rm~ years})}$. $^\dagger$Sources projected onto extinction higher than 0.5 mags do not have any distance/velocity reported}
\end{deluxetable}

\clearpage
%%%%%%%%%%%%%%%%%%%%%%%%%%%%%%%%
%%%%%%%%% FIGURES %%%%%%%%%%%%%%
\onecolumn

%%%%%%%%%%%%%%%%%%%%%%%%%%%%%%%%%%%
%%%%%%%%% APPENDICES %%%%%%%%%%%%%%
\appendix

\renewcommand\thefigure{\thesection}
\renewcommand\thesubfigure{.\arabic{subfigure}}
\setcounter{figure}{0}

\clearpage

\section{Images of Confirmed Protostars}\label{images}

Images of \stageI\ confirmed protostars and source \M1246, which is included to show it's jet. On the left is the VISION Survey JHK $1\arcmin\times1\arcmin$ image. The green circle marks the source, and the red circles mark Megeath et al. Class II sources. On the right is the Herschel $30\arcsec$ resolution $10\arcmin\times10\arcmin$ extinction image. $\ak$ ranges from $\ak$ of 0-0.8 mag. The blue square is a $1\arcmin\times1\arcmin$ box centered on the source. Green circles mark the Megeath et al. Class 0/I sources; red X's, the Megeath et al. Class II sources. In all extinction images, the colorbar goes from white (0) to black (0.8) magnitudes in $\ak$. For the sources with jets, the jet is marked with an arrow. Because jet's should be bipolar, we show the undetected reverse jet as a dashed arrow. Orientation is galactic north up - east left
\begin{figure}
\centering
\subfloat{\label{fig:0333}\source{0333}} \\
\subfloat{\label{fig:1246}\source{1246}} \\
\subfloat{\label{fig:1316}\source{1316}}
\caption{\scriptsize\label{fig:images}}
\end{figure}

\begin{figure}
\ContinuedFloat
\subfloat{\label{fig:1369}\source{1369}} \\
\subfloat{\label{fig:1512}\source{1512}} \\
\subfloat{\label{fig:2393}\source{2393}}
\caption{continued}
\end{figure}

\begin{figure}
\ContinuedFloat
\subfloat{\label{fig:2593}\source{2593}} \\
\subfloat{\label{fig:2649}\source{2649}} \\
\subfloat{\label{fig:2690}\source{2690}}
\caption{continued}
\end{figure}

\begin{figure}
\ContinuedFloat
\subfloat{\label{fig:2707}\source{2707}} \\
\subfloat{\label{fig:2748}\source{2748}}
\caption{continued}
\end{figure}

\clearpage

\section{SED Atlas}\label{atlas}
This appendix shows the model fits and data for all the Megeath protostars below $\ak$ of 0.8 mag except for the \tld\ that had too little data. The gray lines are the fits with $\Delta\chi^2_{scaled} \le 1$, with the best fit in black. The IRS spectrum, where available, is overplotted in red.
\begin{center}
\begin{longtable}{ccc}
\ppcell{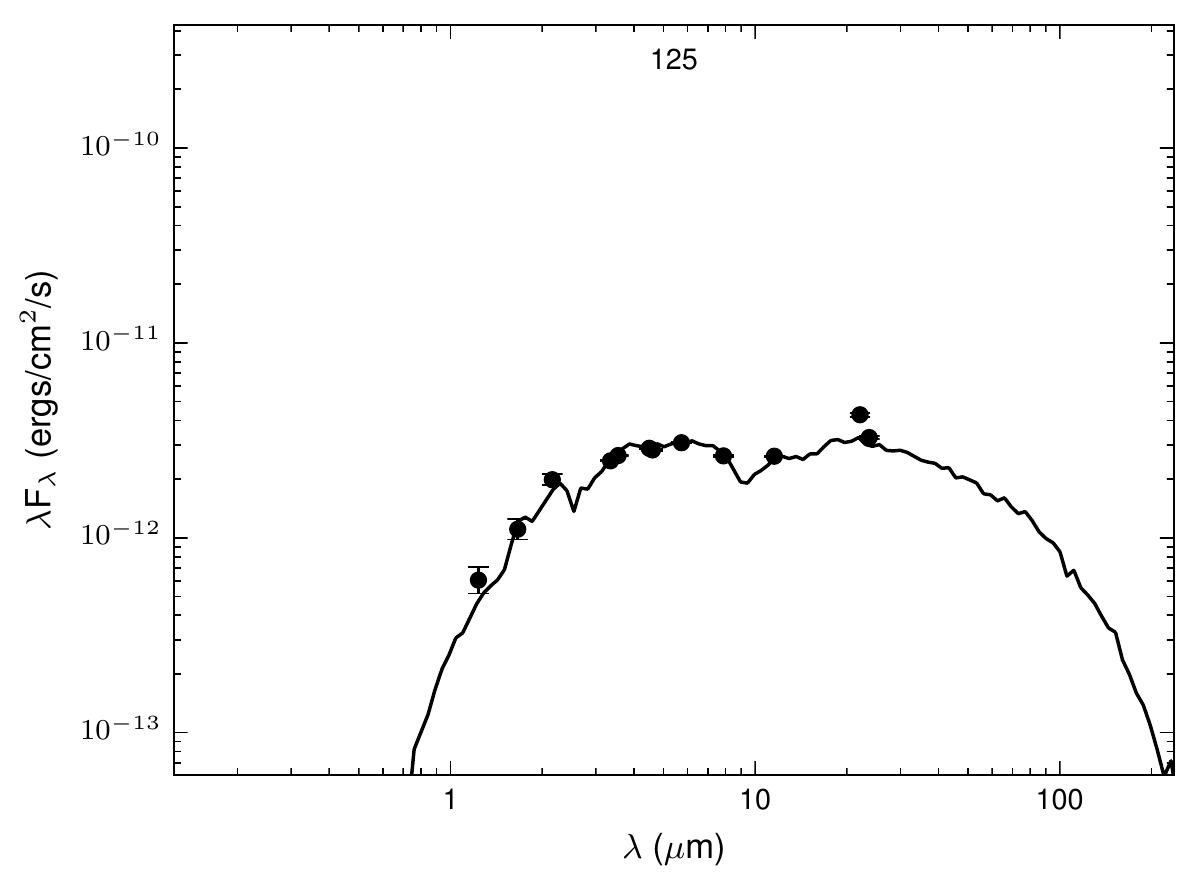} & \ppcell{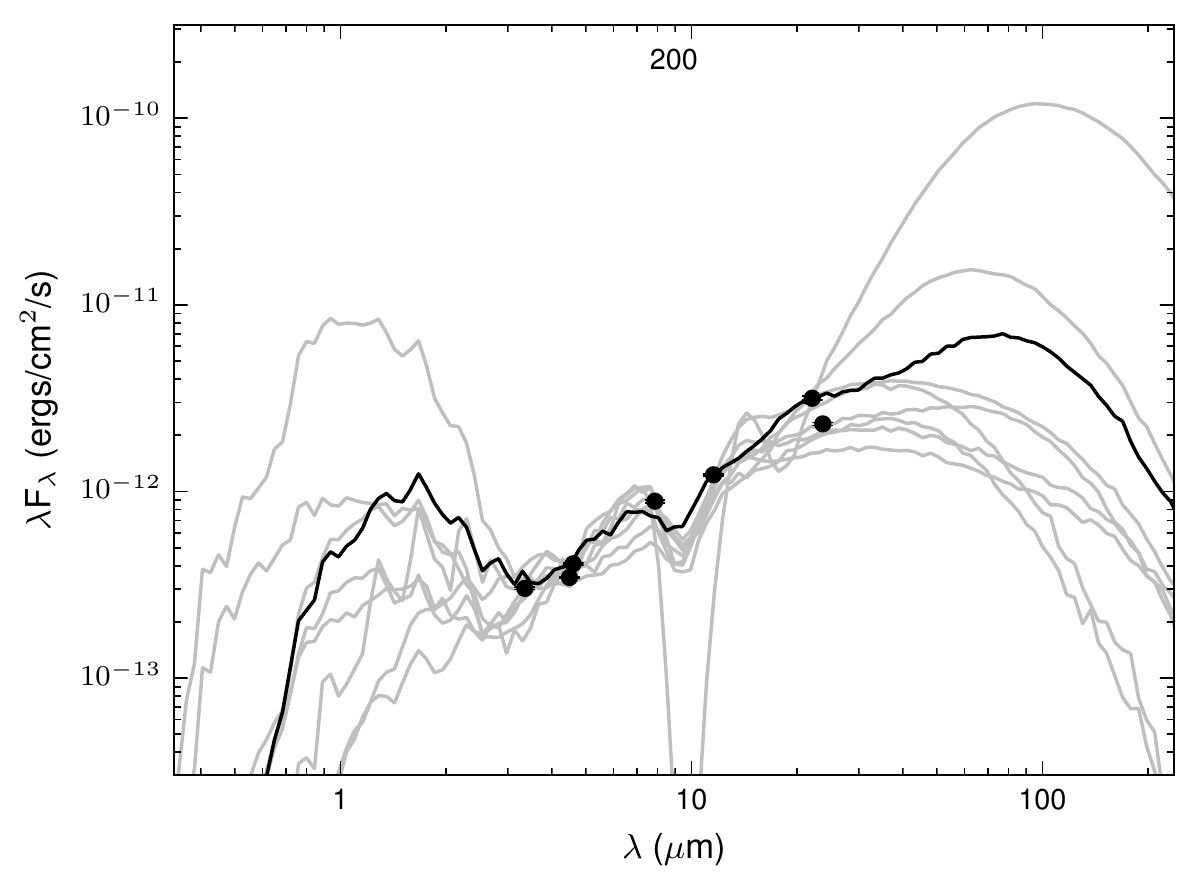} & \ppcell{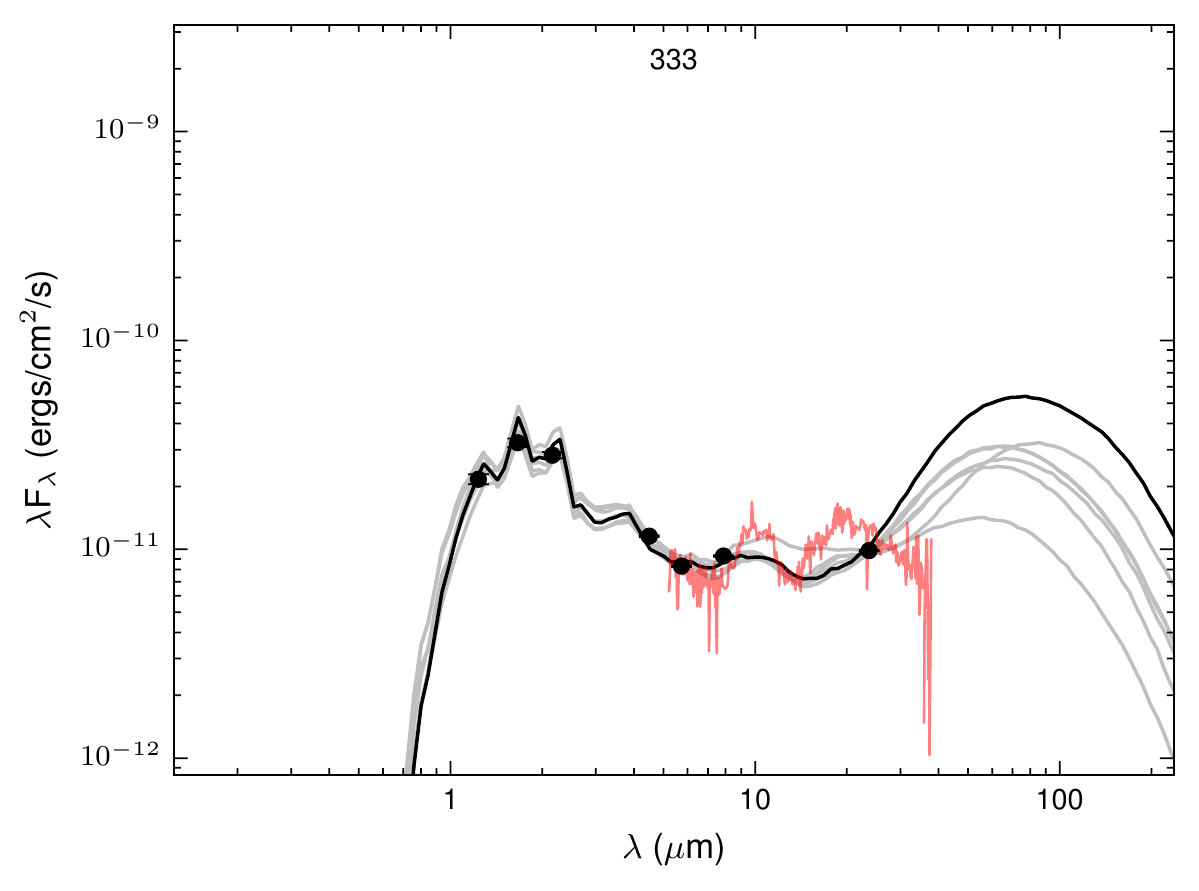} \\
\ppcell{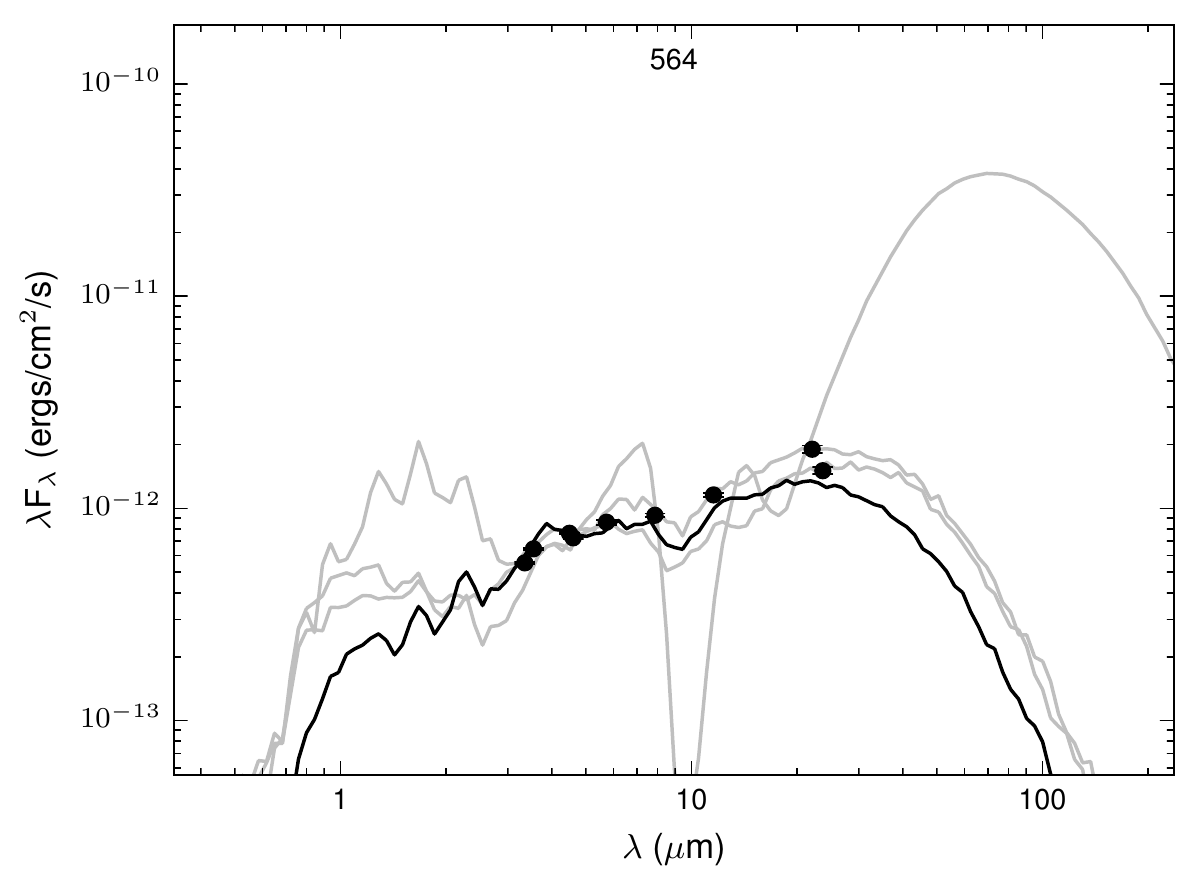} & \ppcell{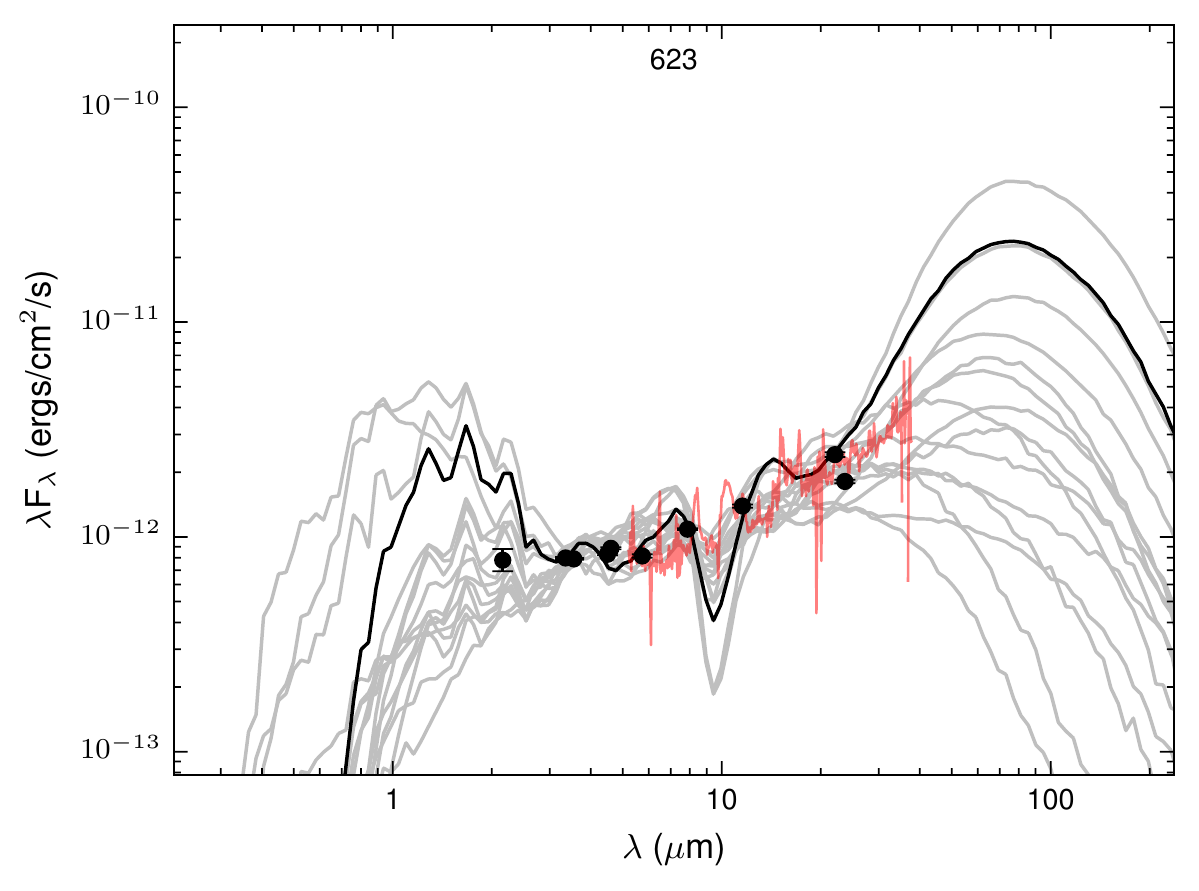} & \ppcell{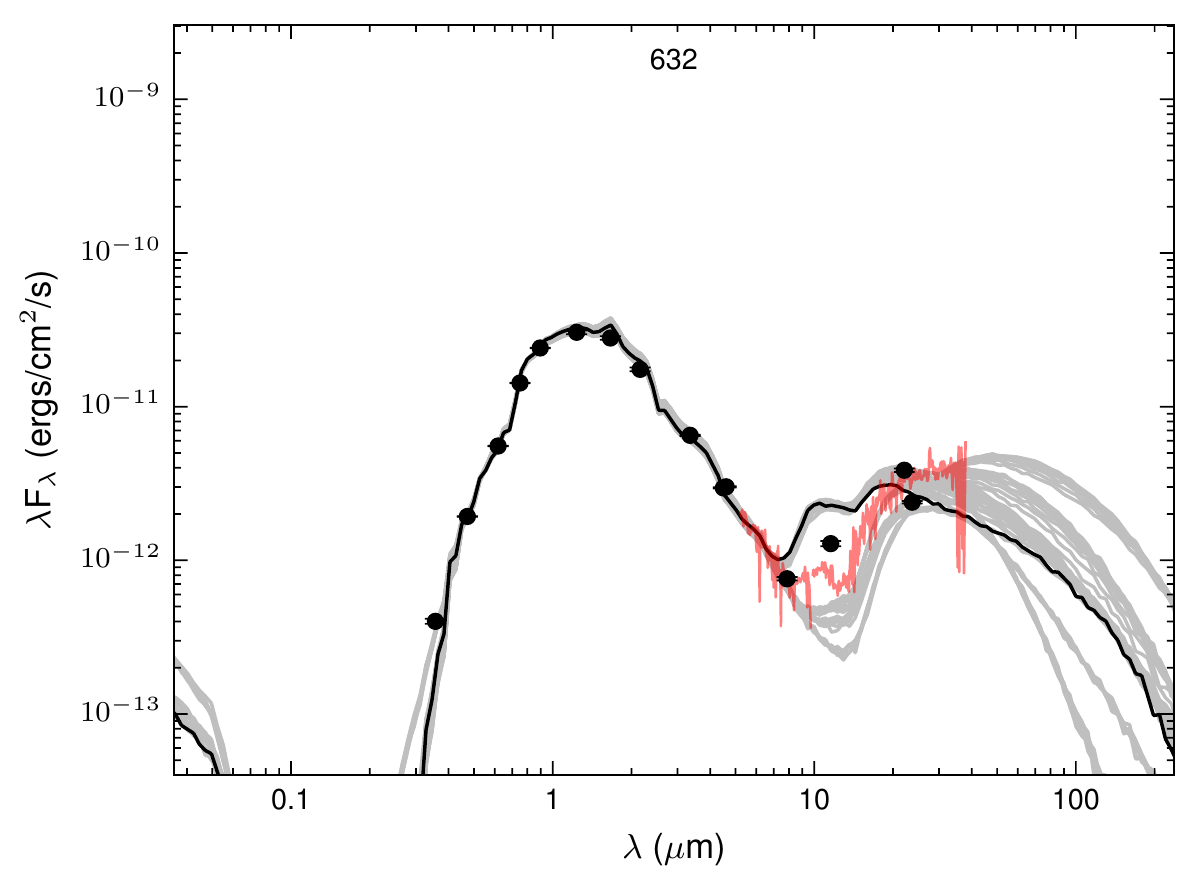} \\
\ppcell{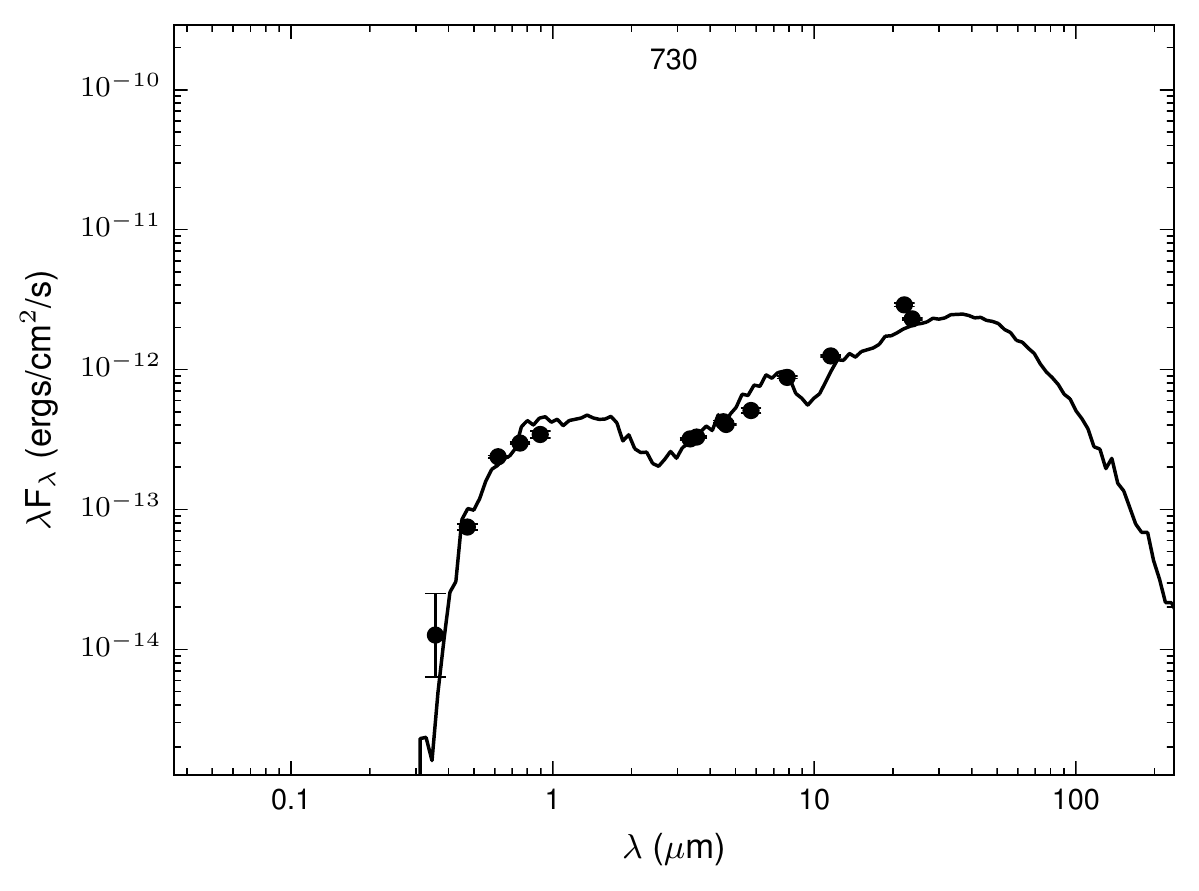} & \ppcell{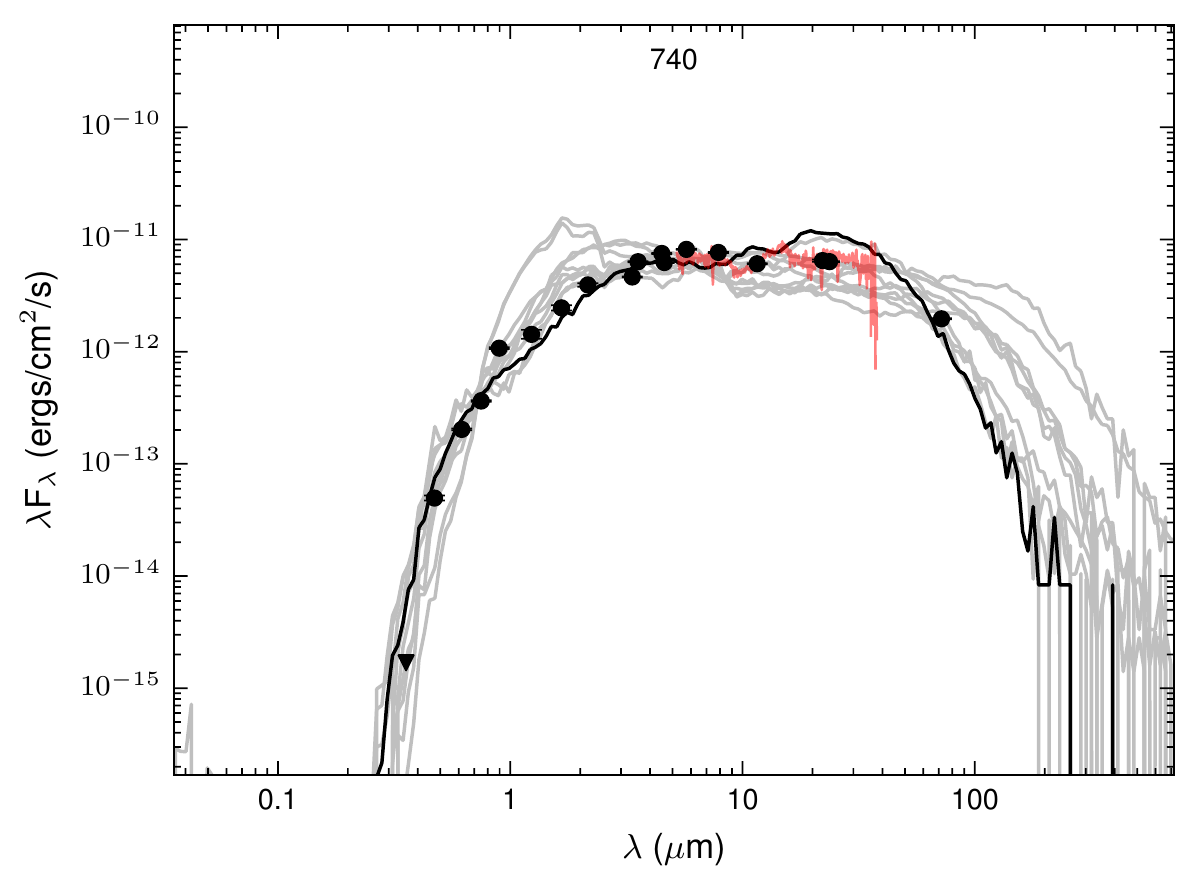} & \ppcell{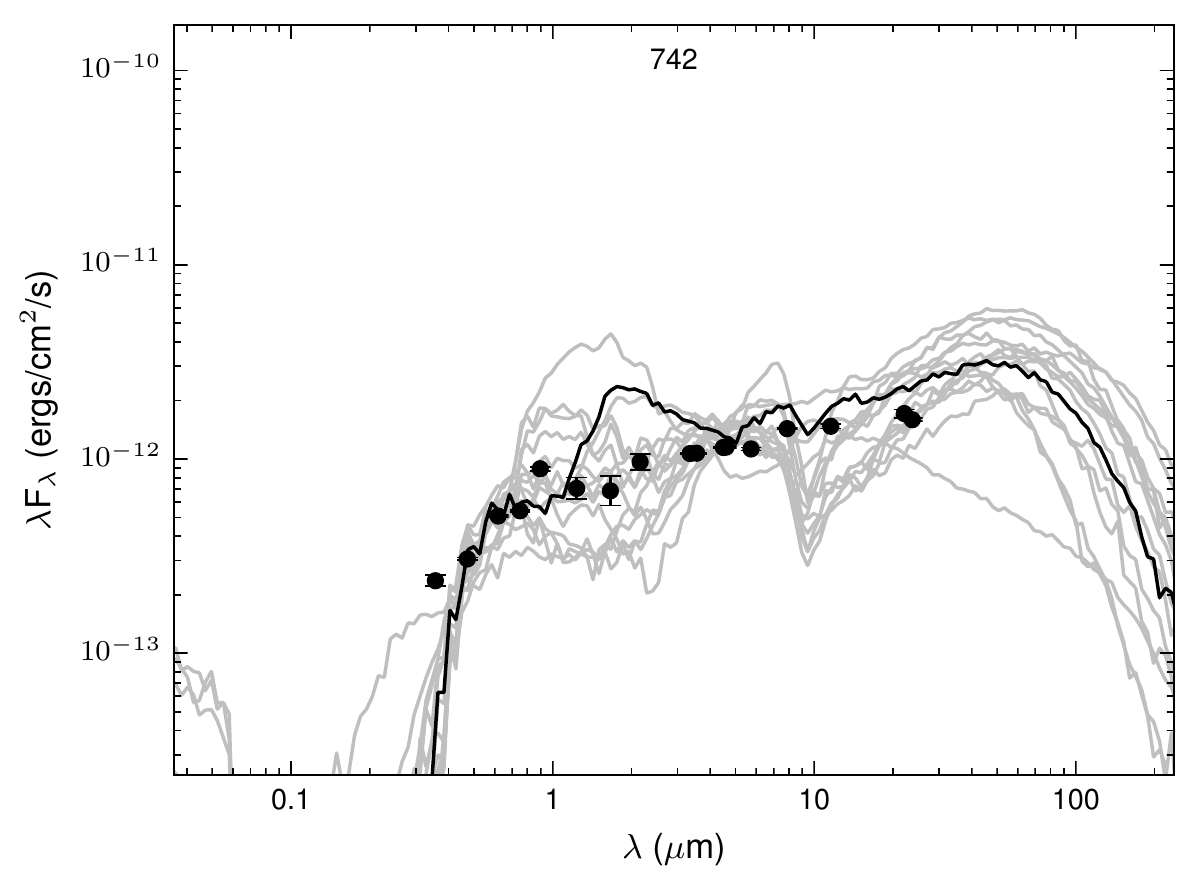} \\
\ppcell{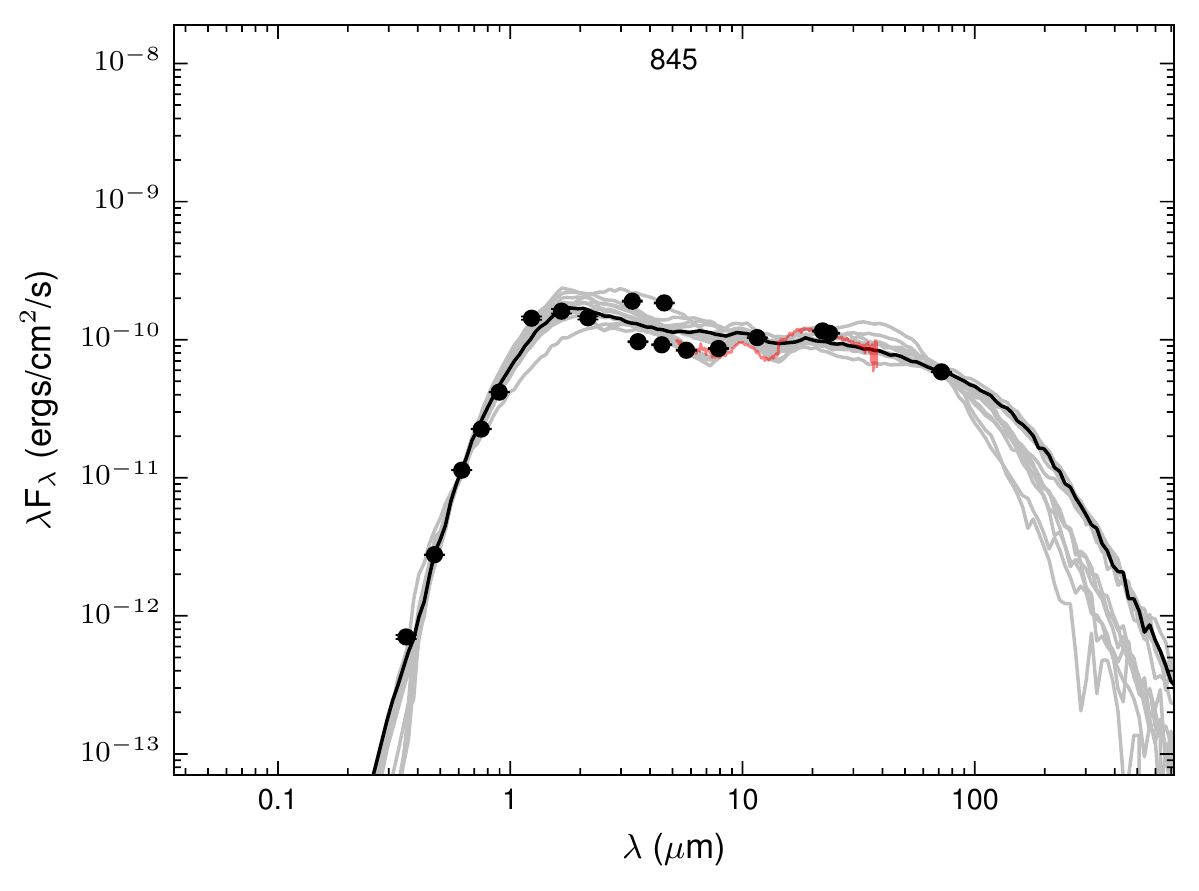} & \ppcell{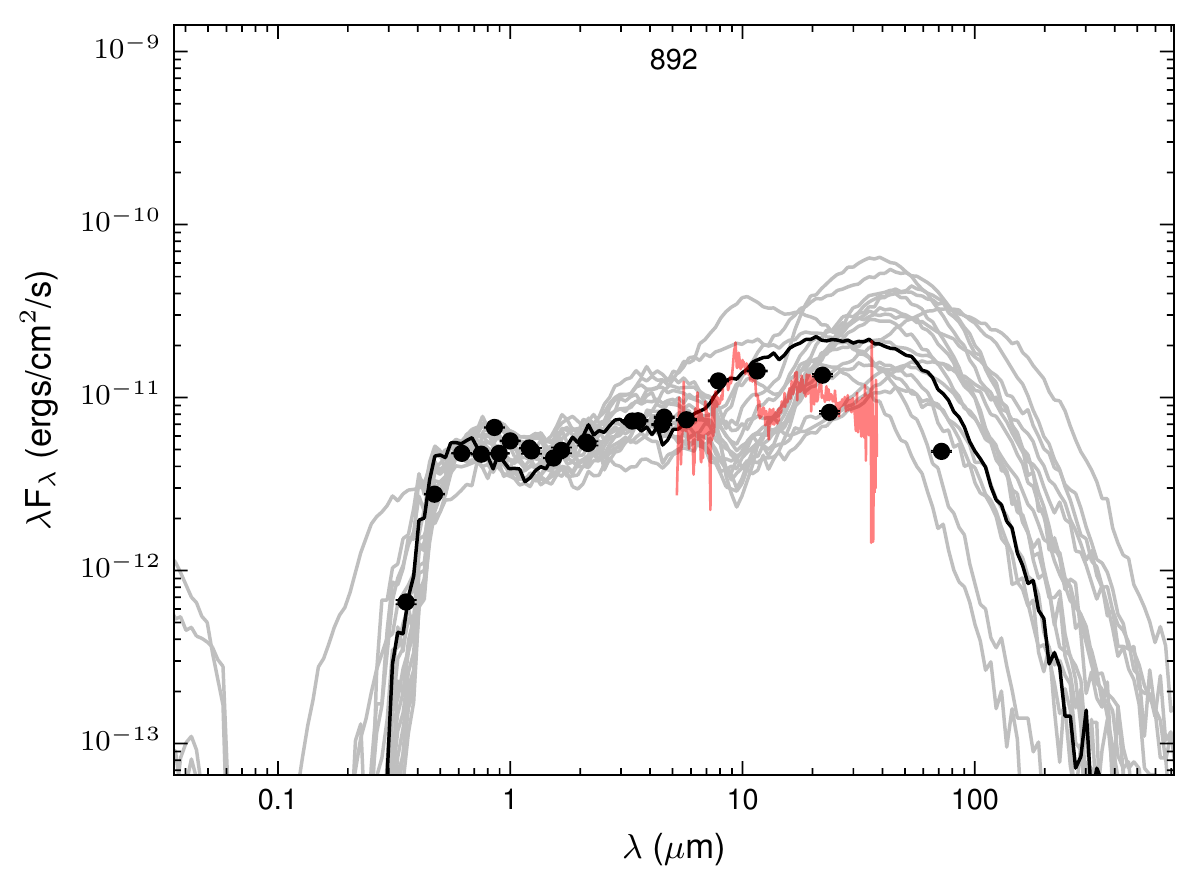} & \ppcell{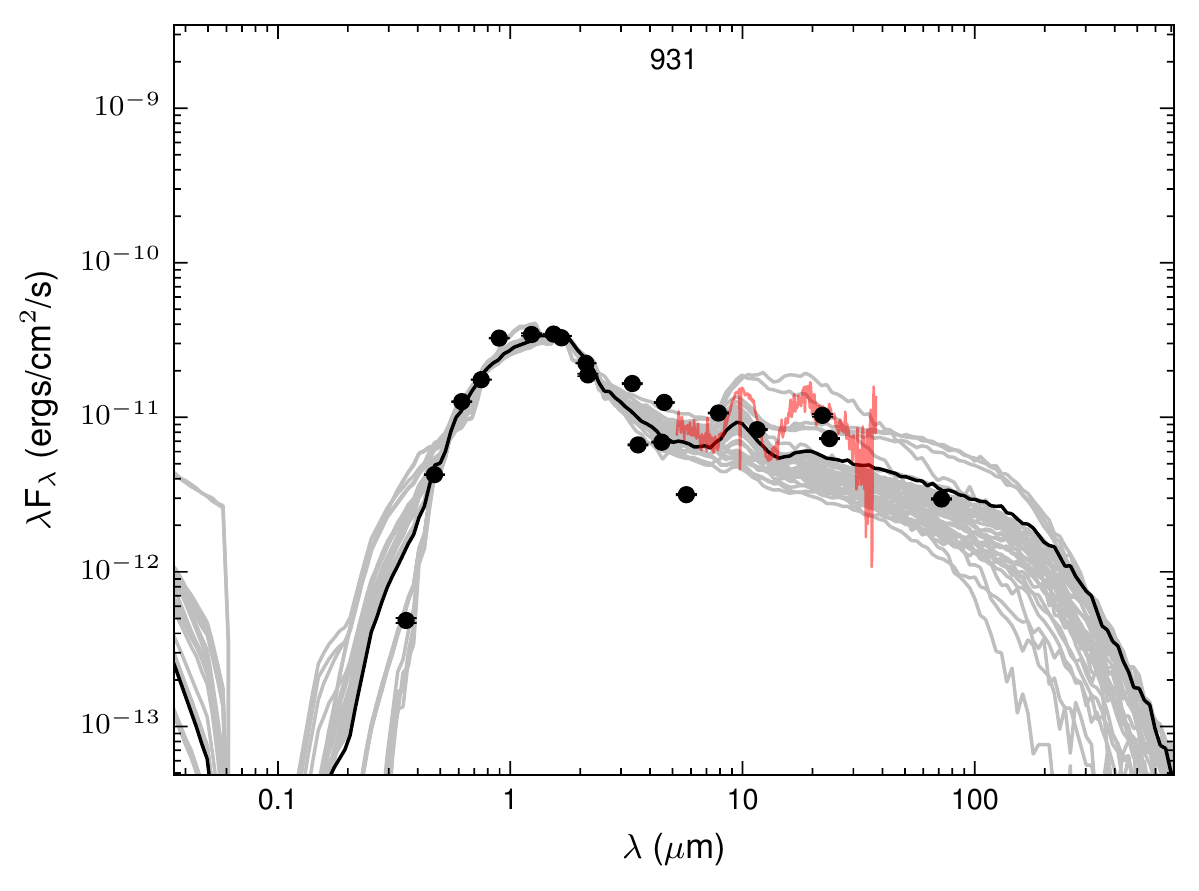} \\
\ppcell{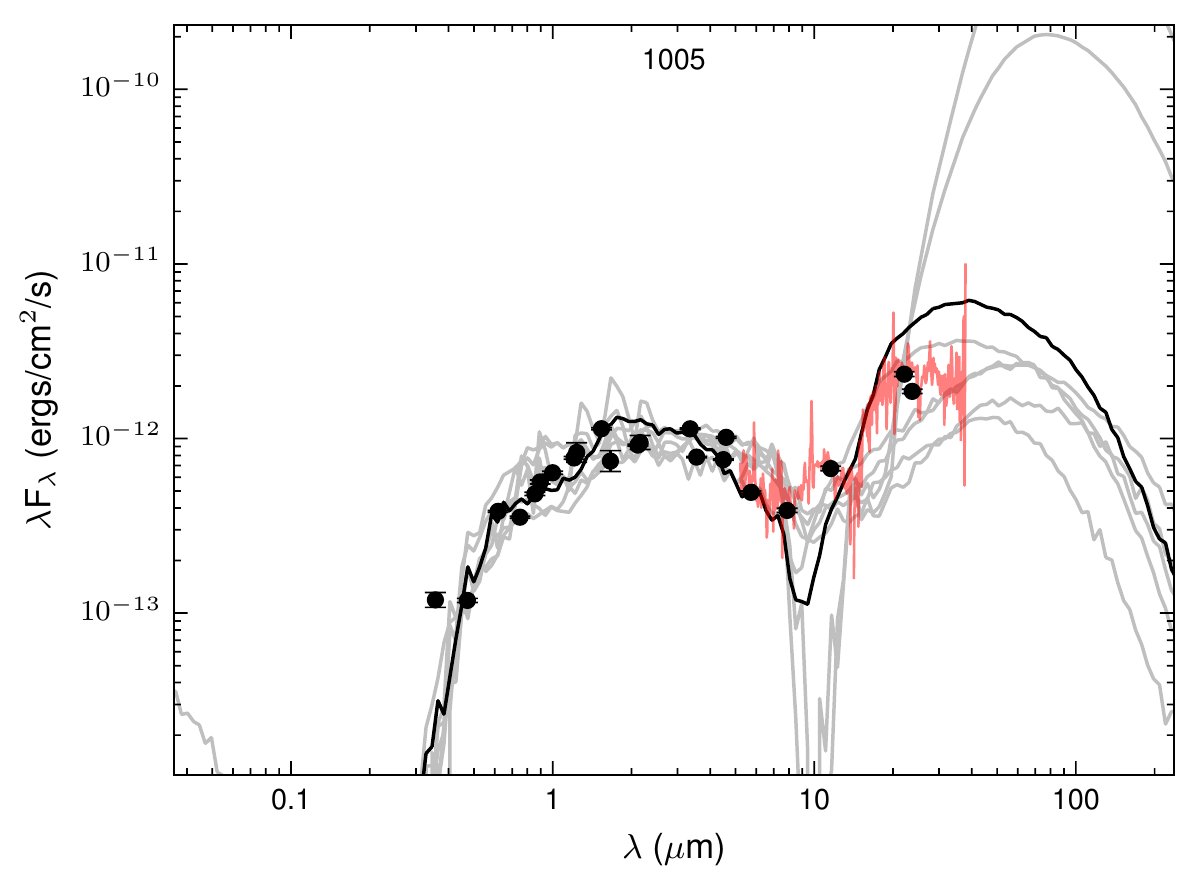} & \ppcell{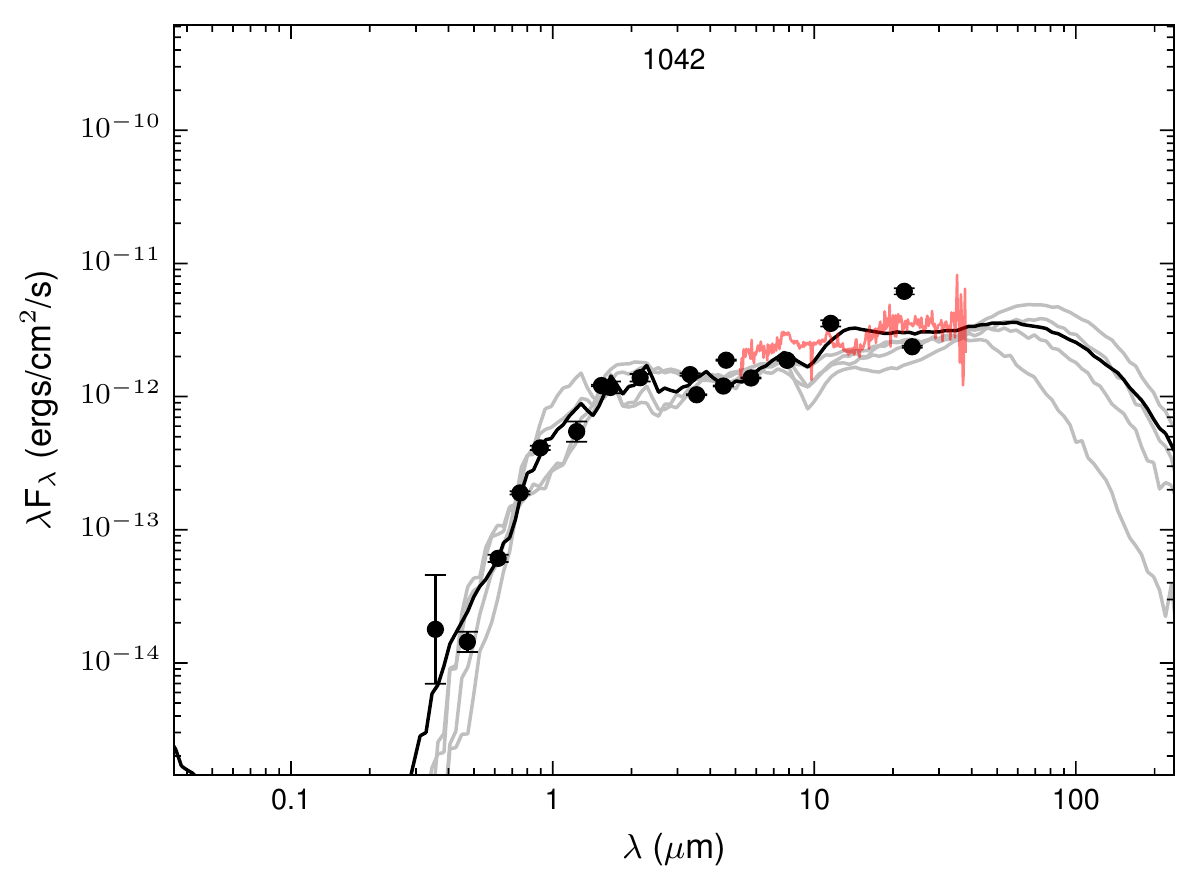} & \ppcell{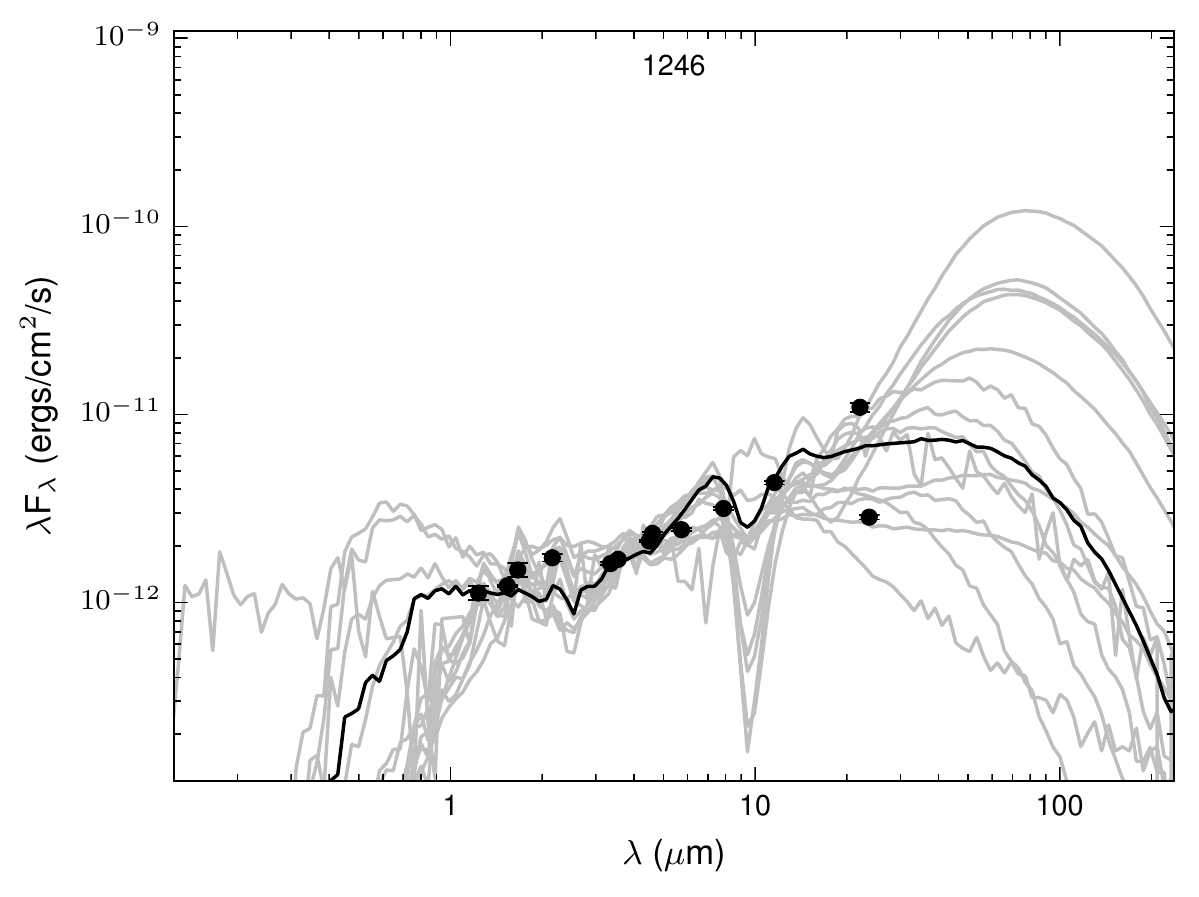} \\
\ppcell{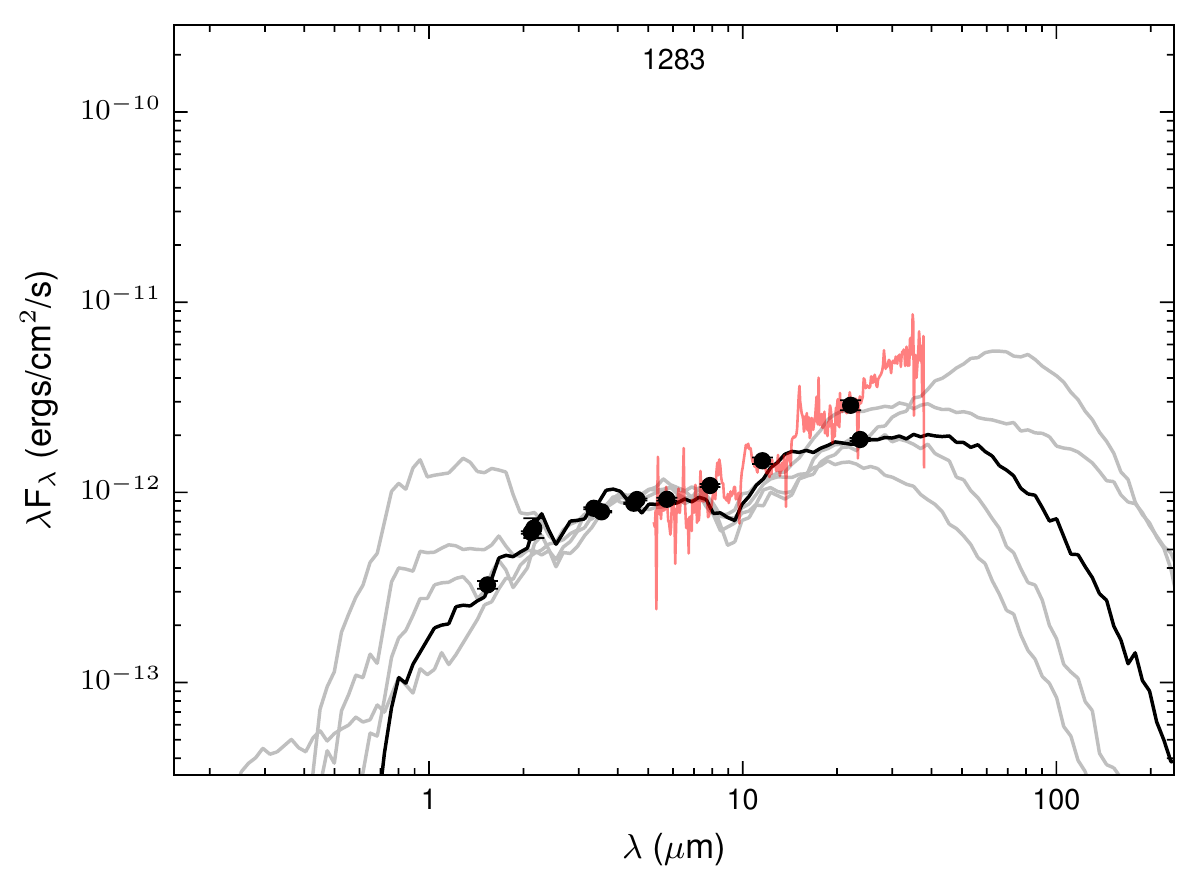} & \ppcell{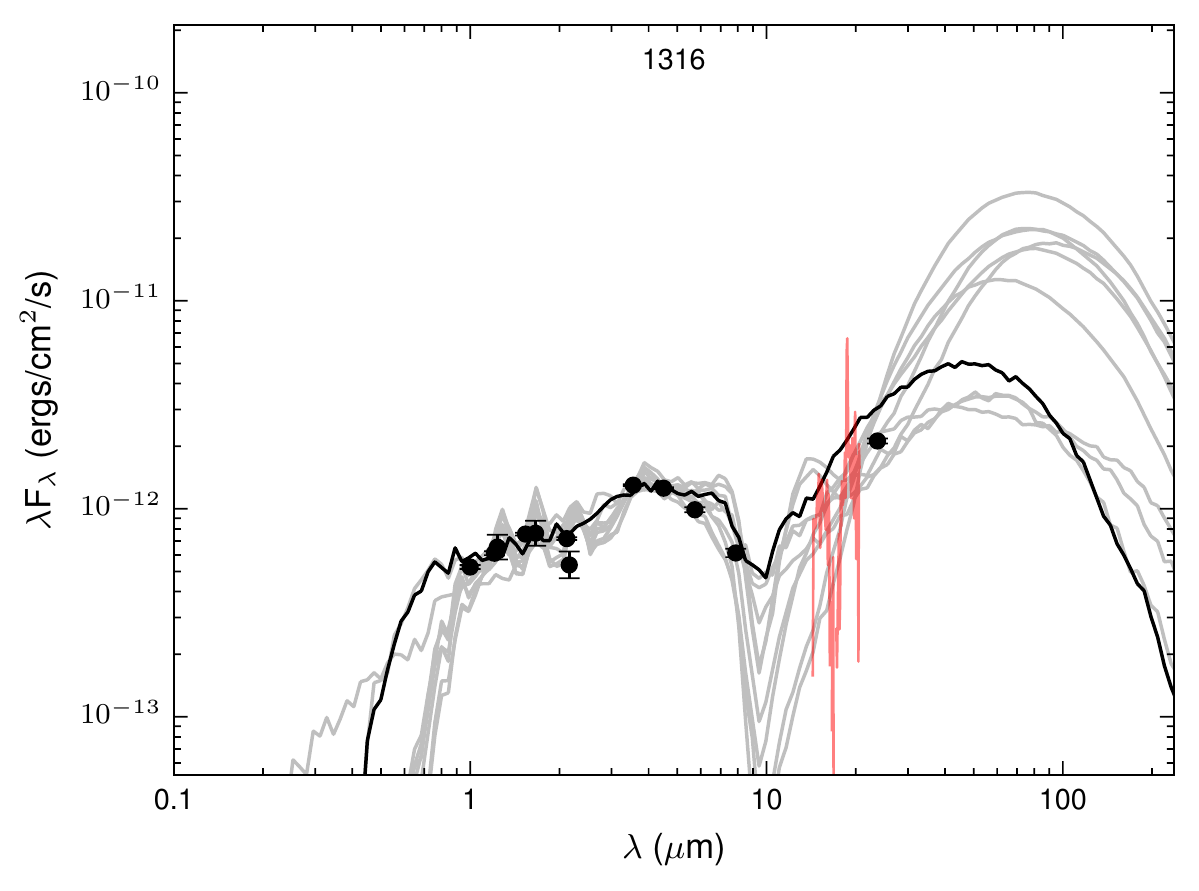} & \ppcell{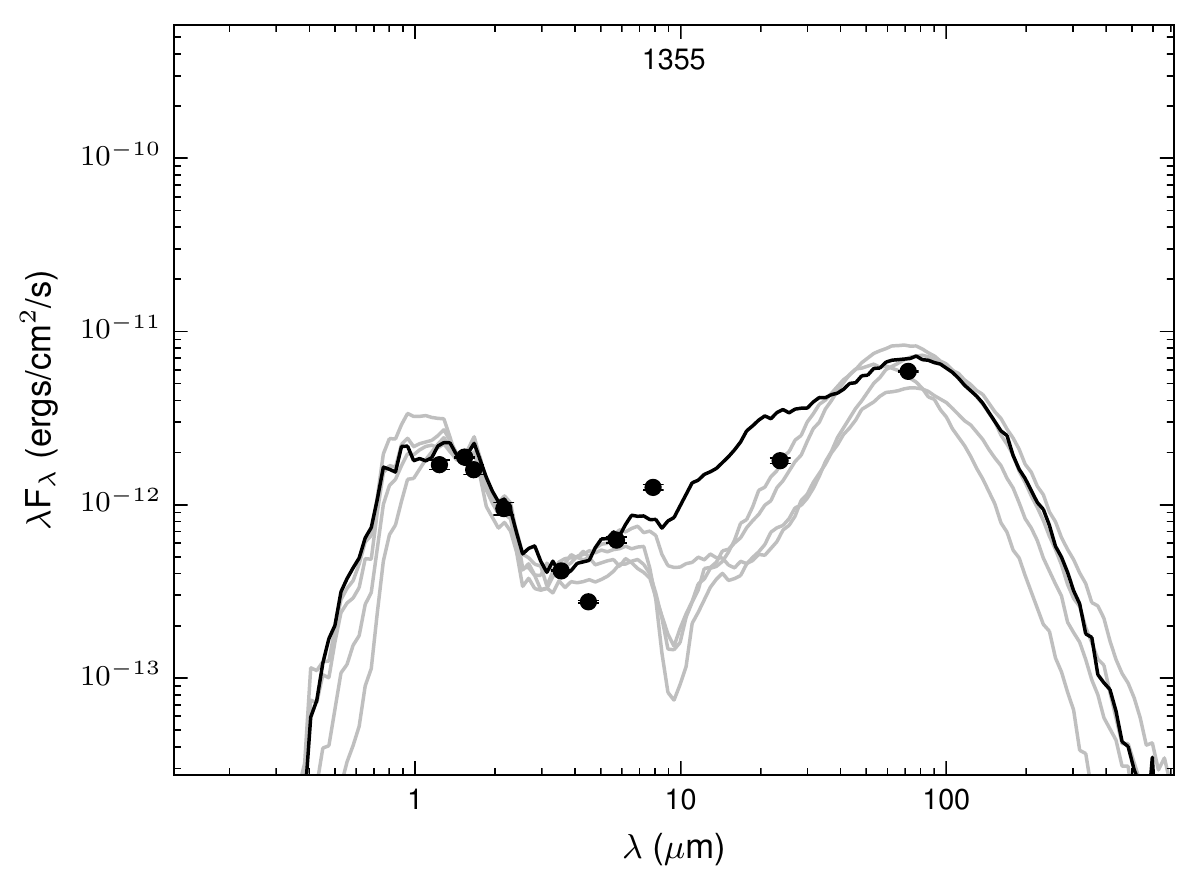} \\
\ppcell{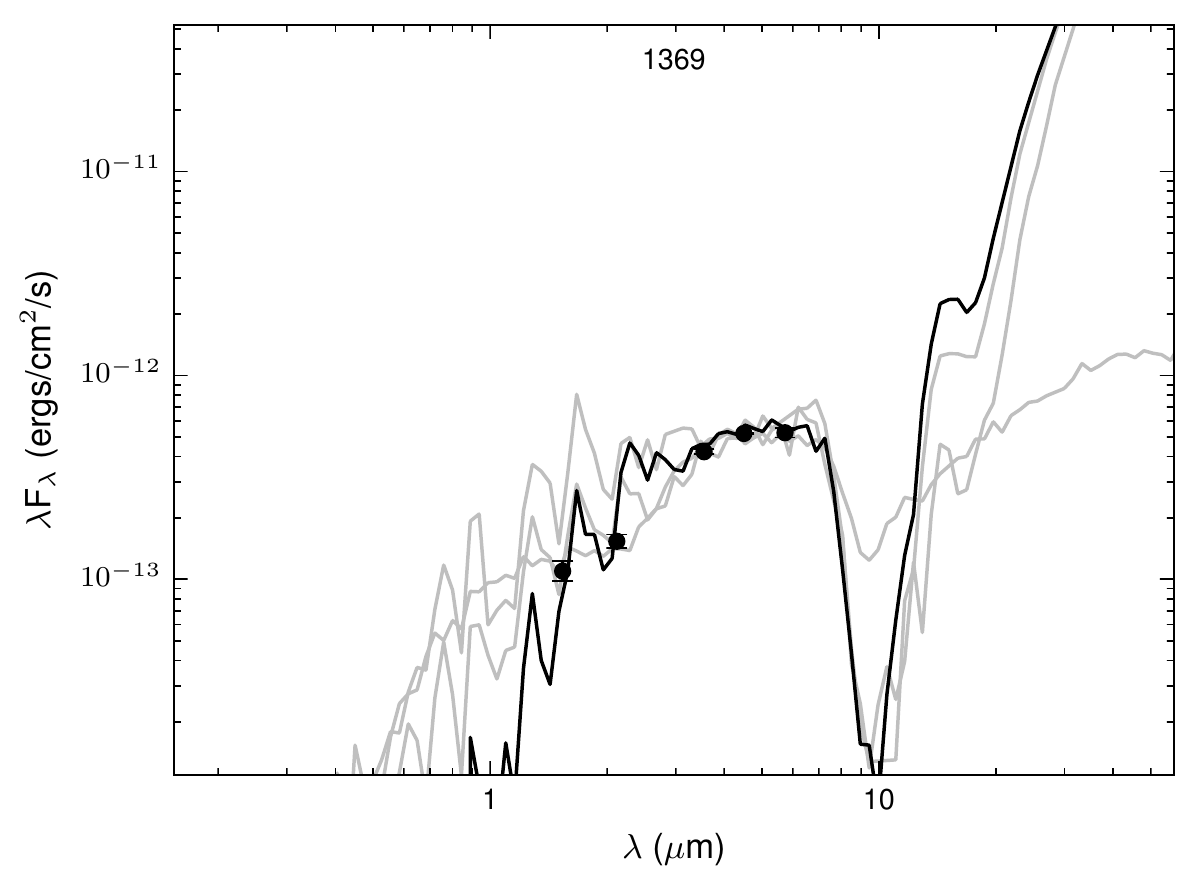} & \ppcell{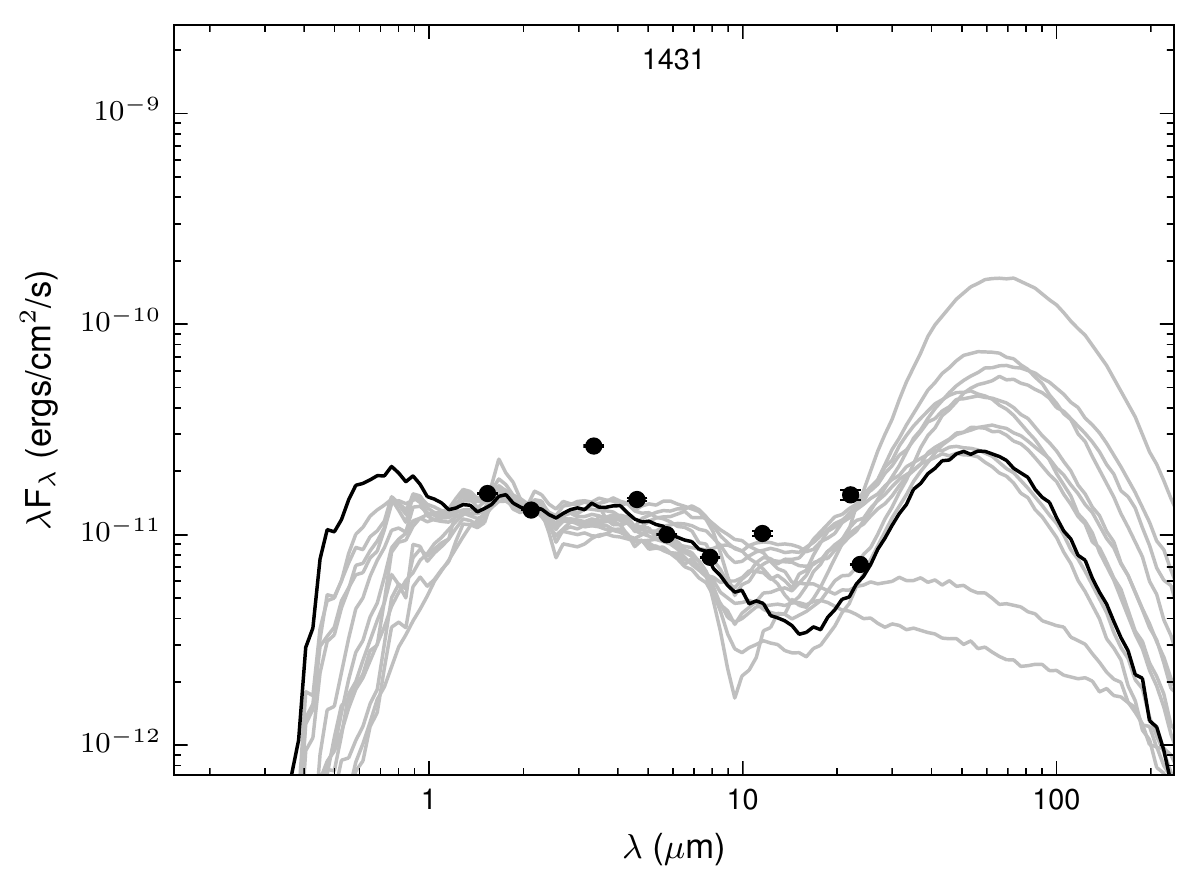} & \ppcell{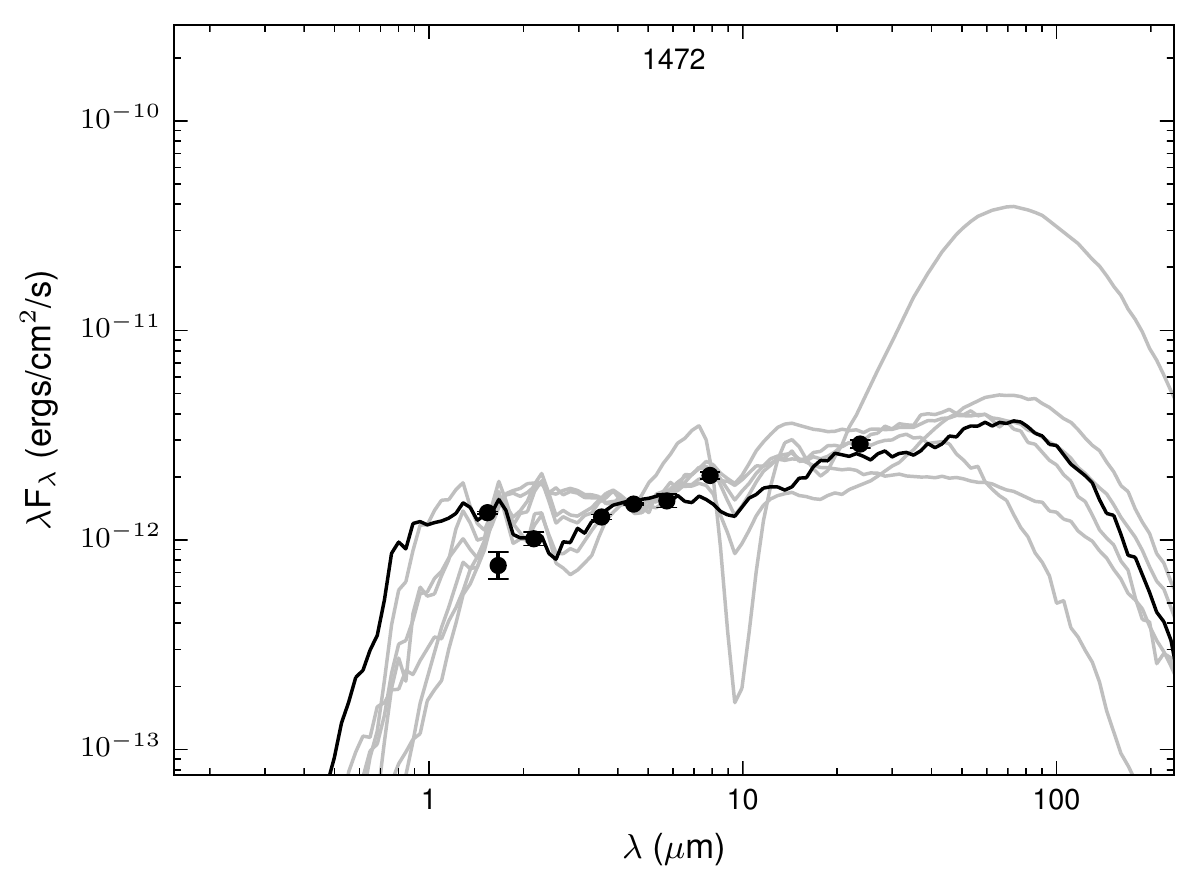} \\
\ppcell{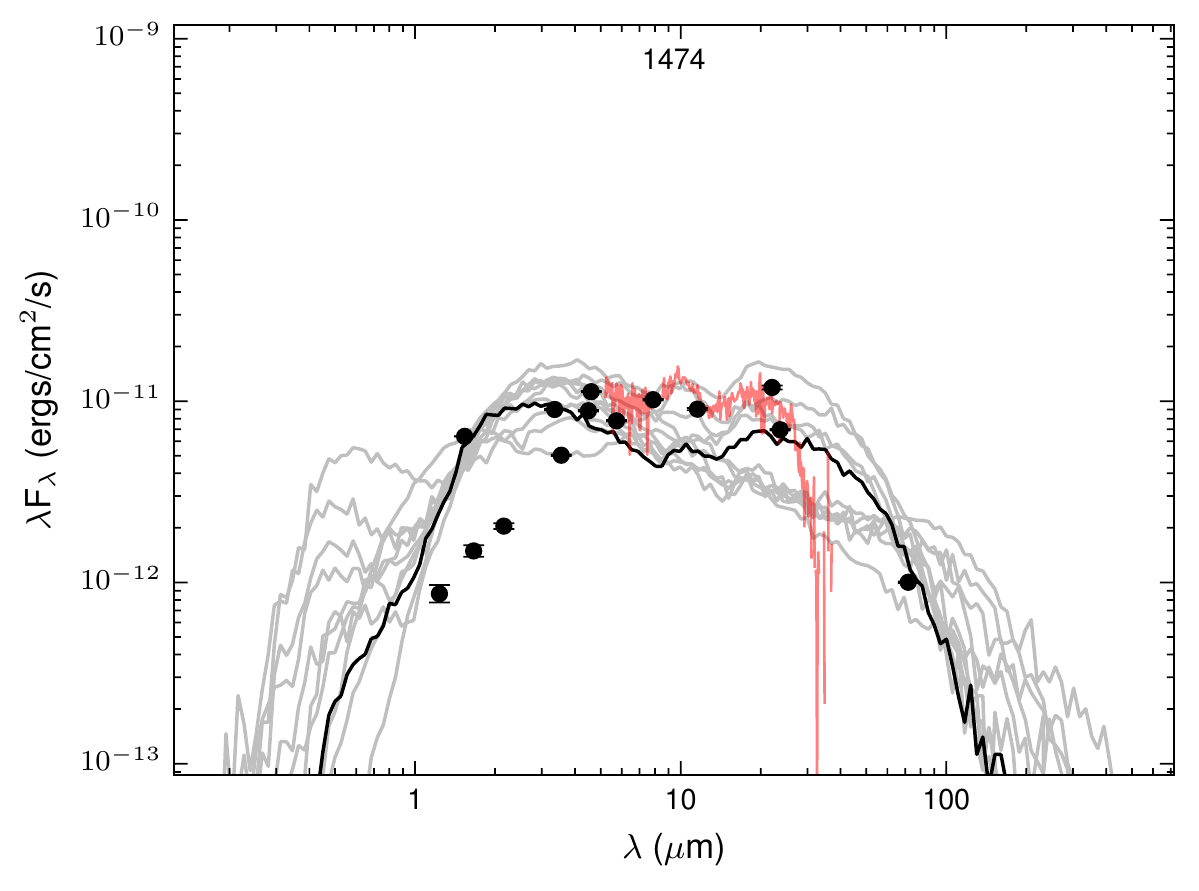} & \ppcell{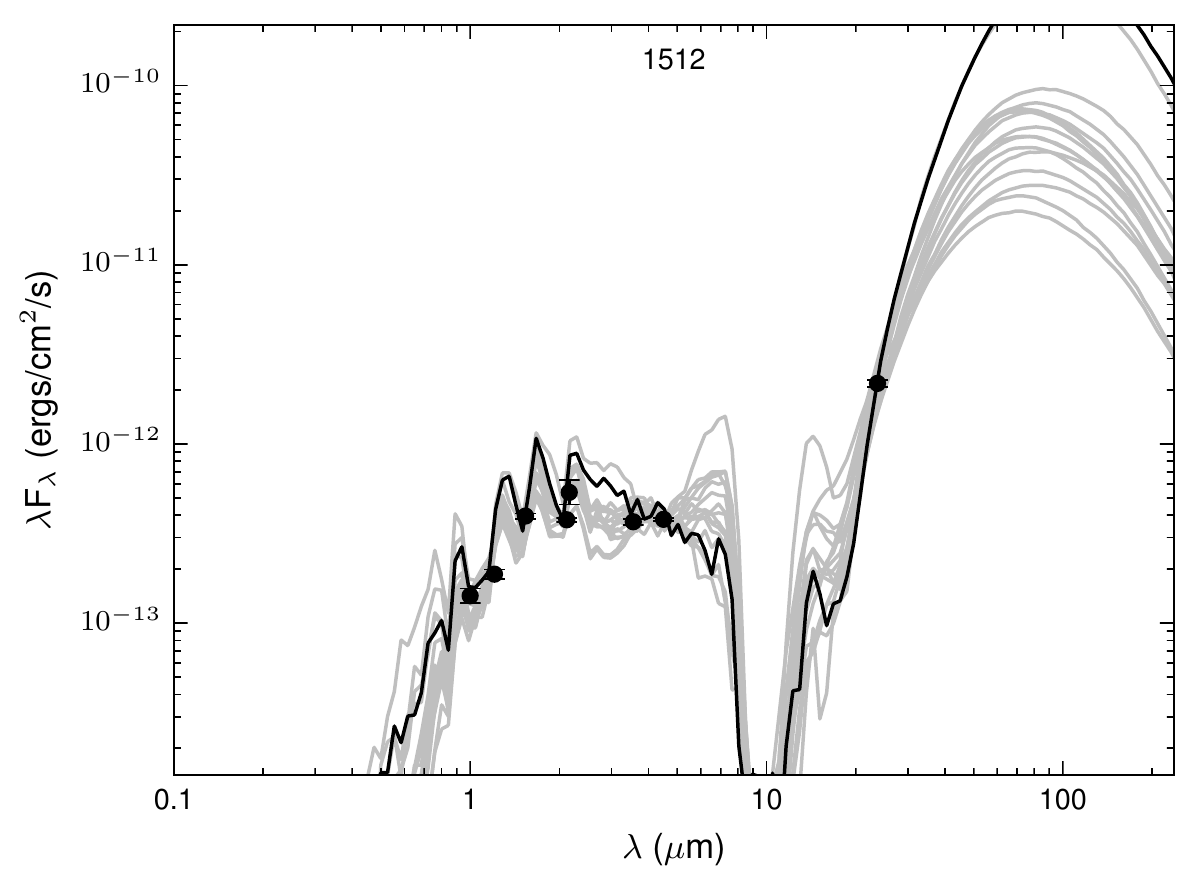} & \ppcell{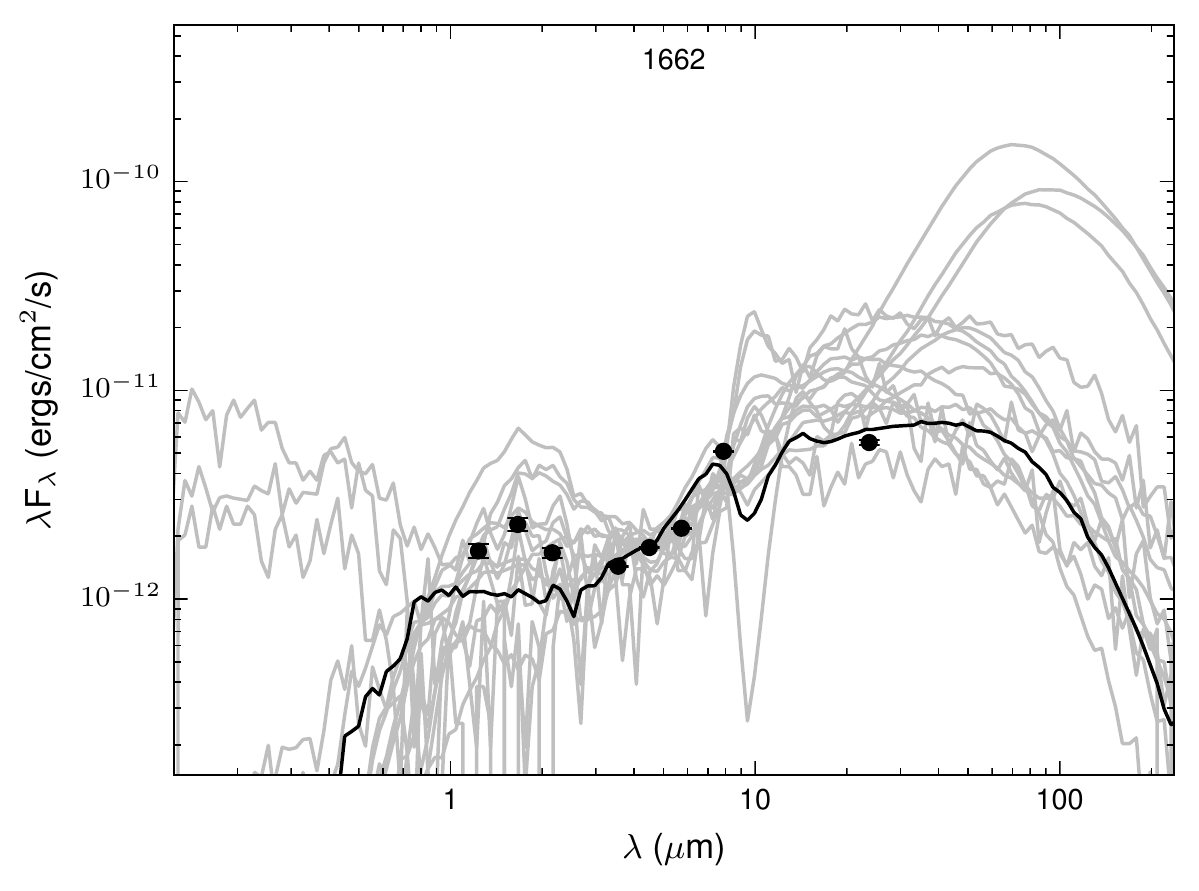} \\
\ppcell{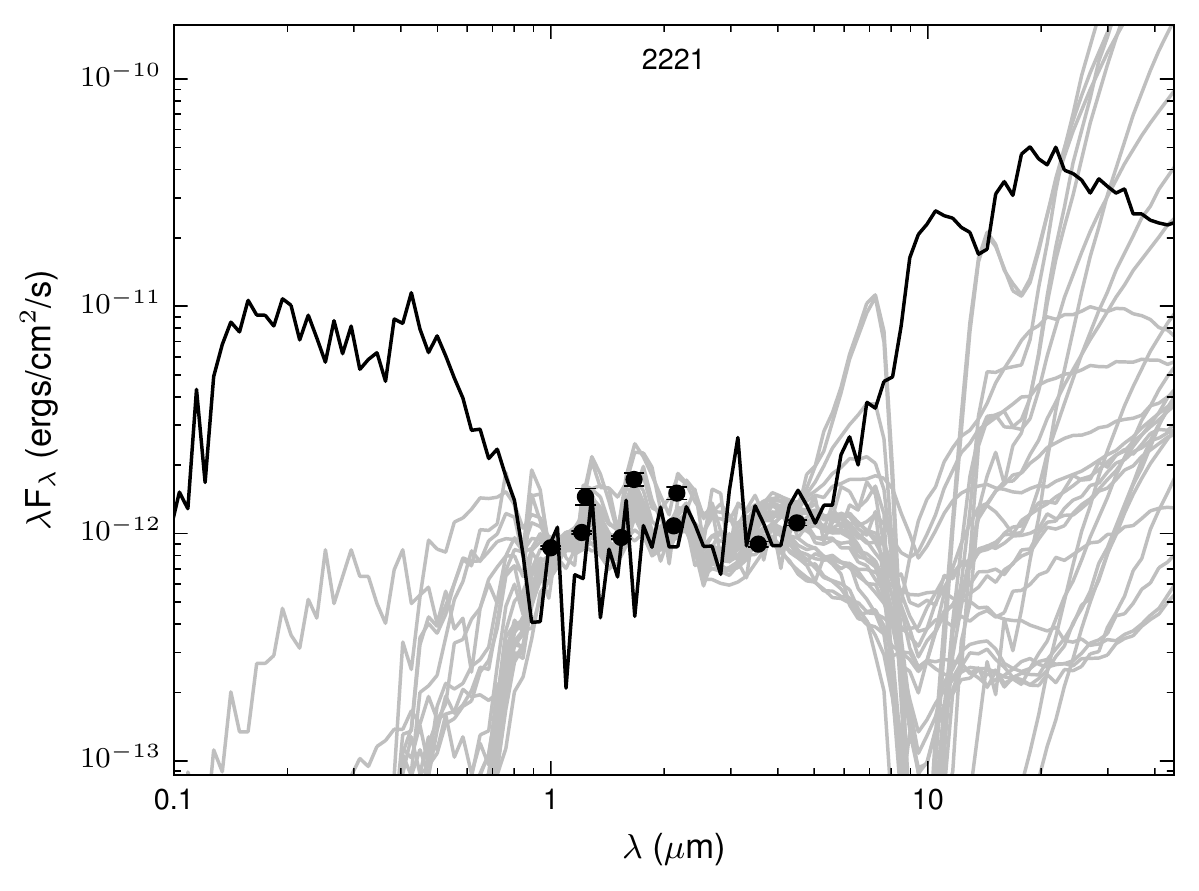} & \ppcell{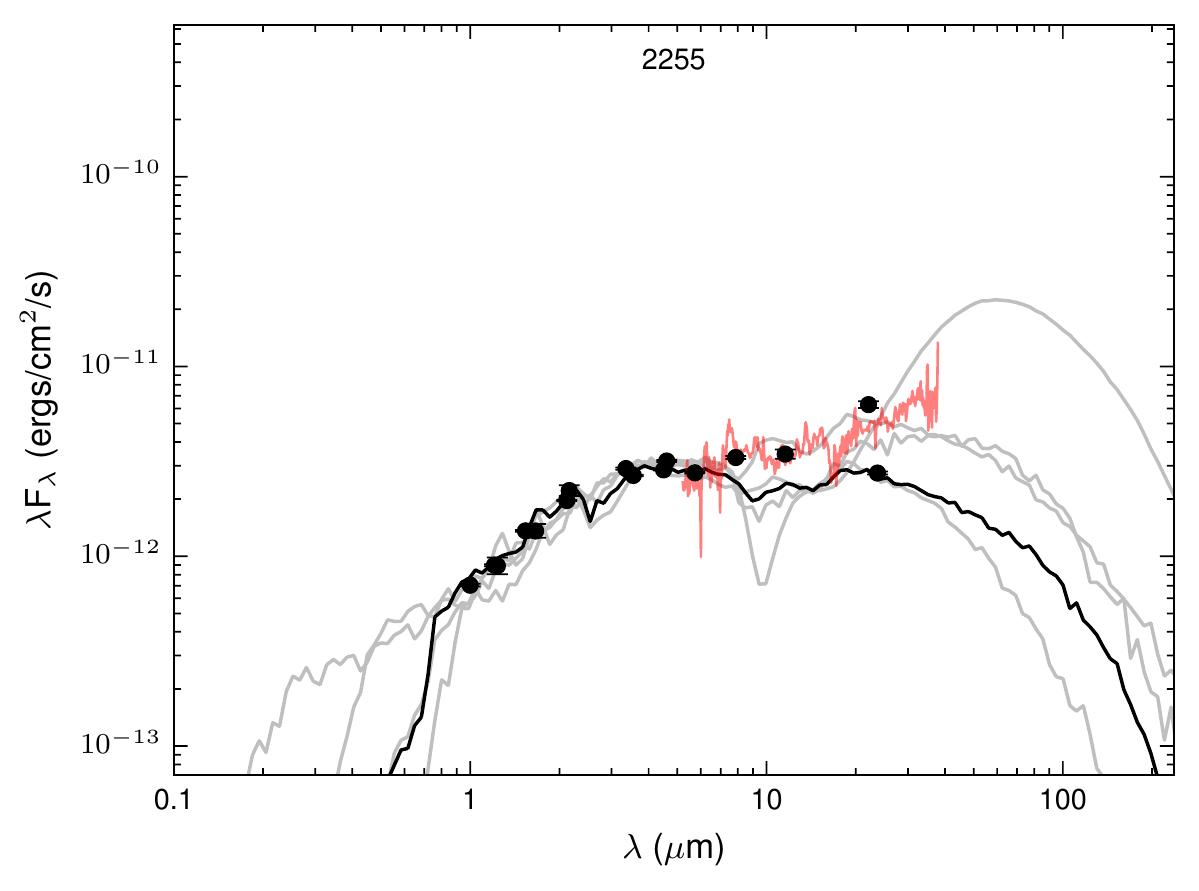} & \ppcell{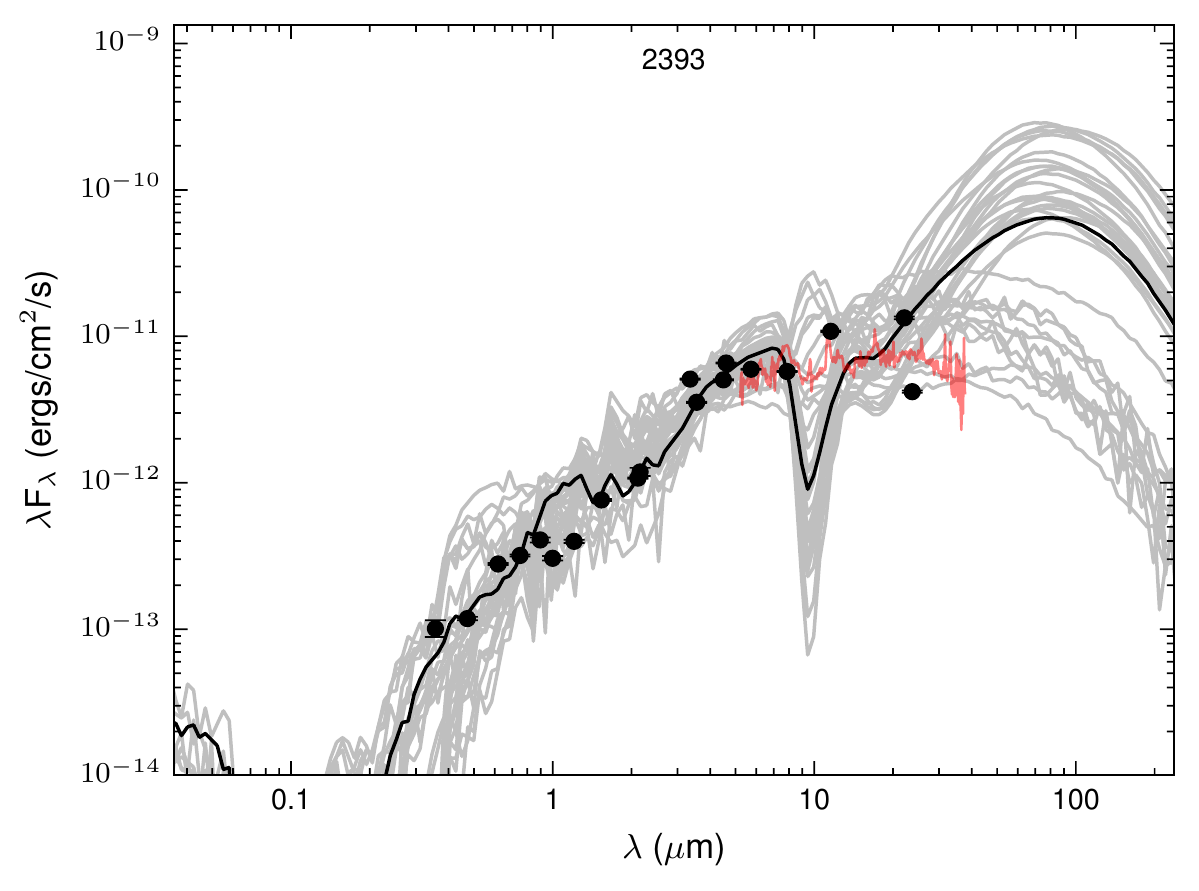} \\
\ppcell{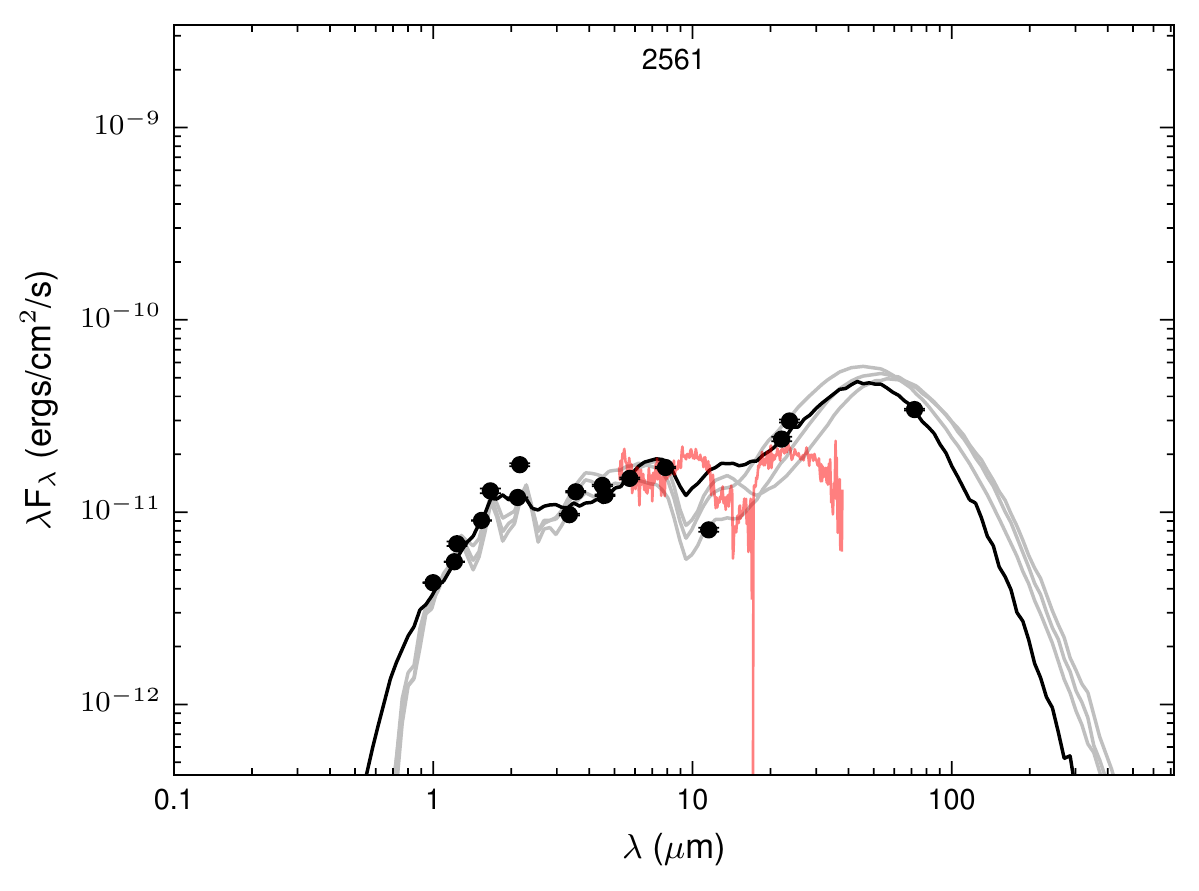} & \ppcell{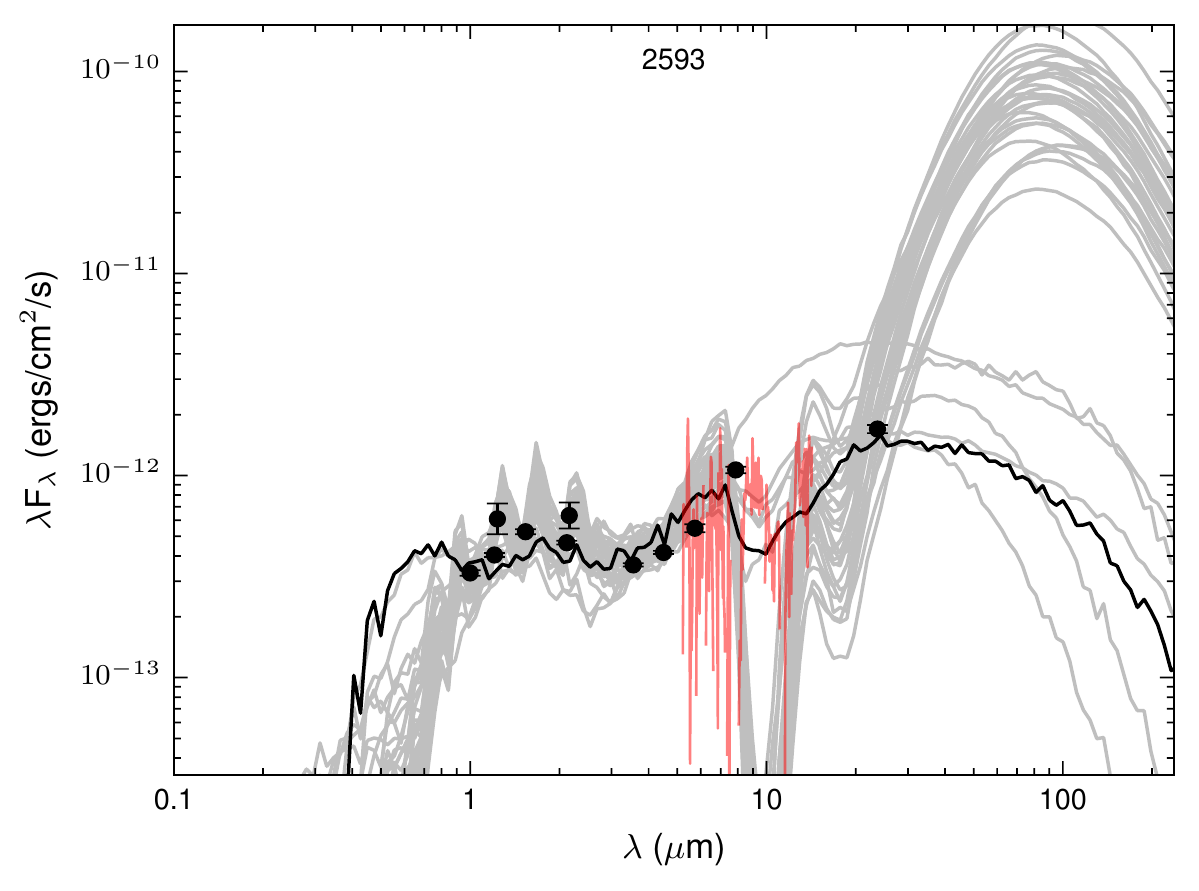} & \ppcell{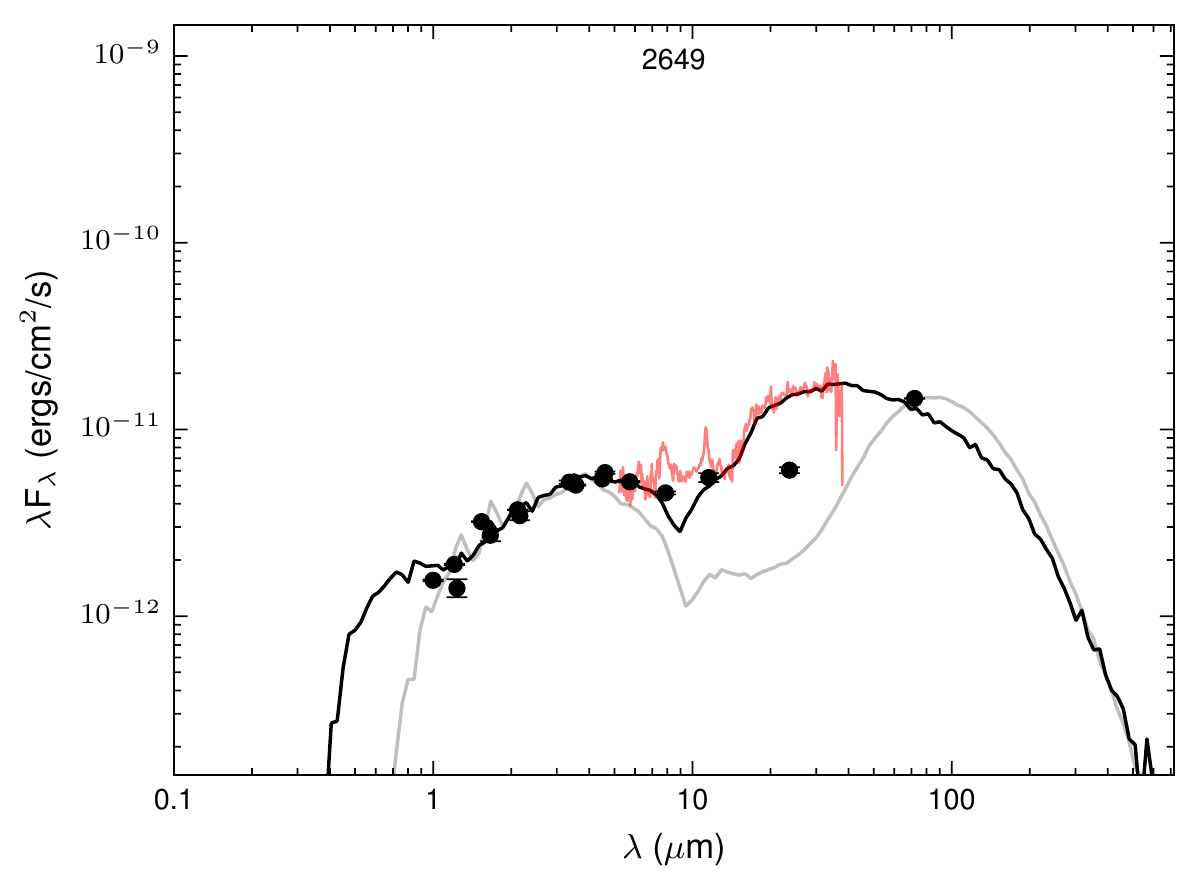} \\
\ppcell{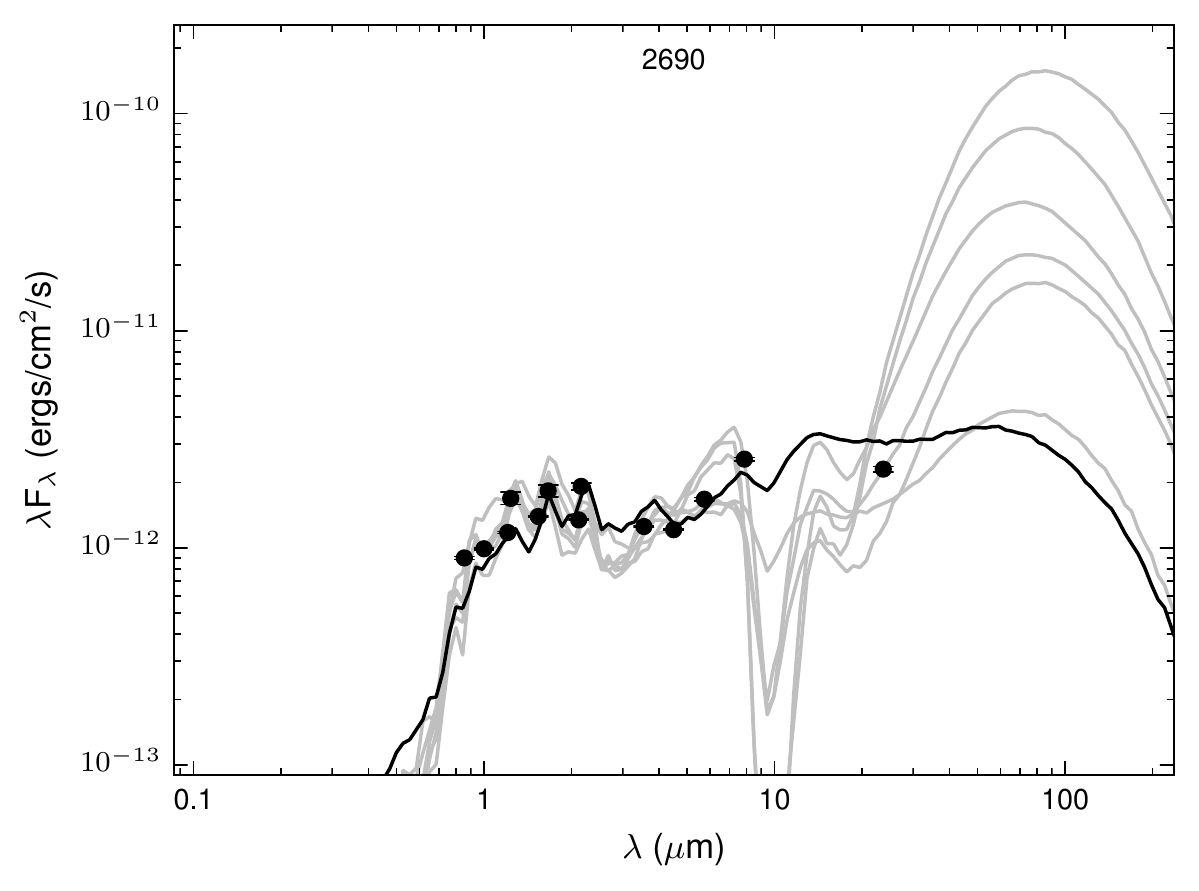} & \ppcell{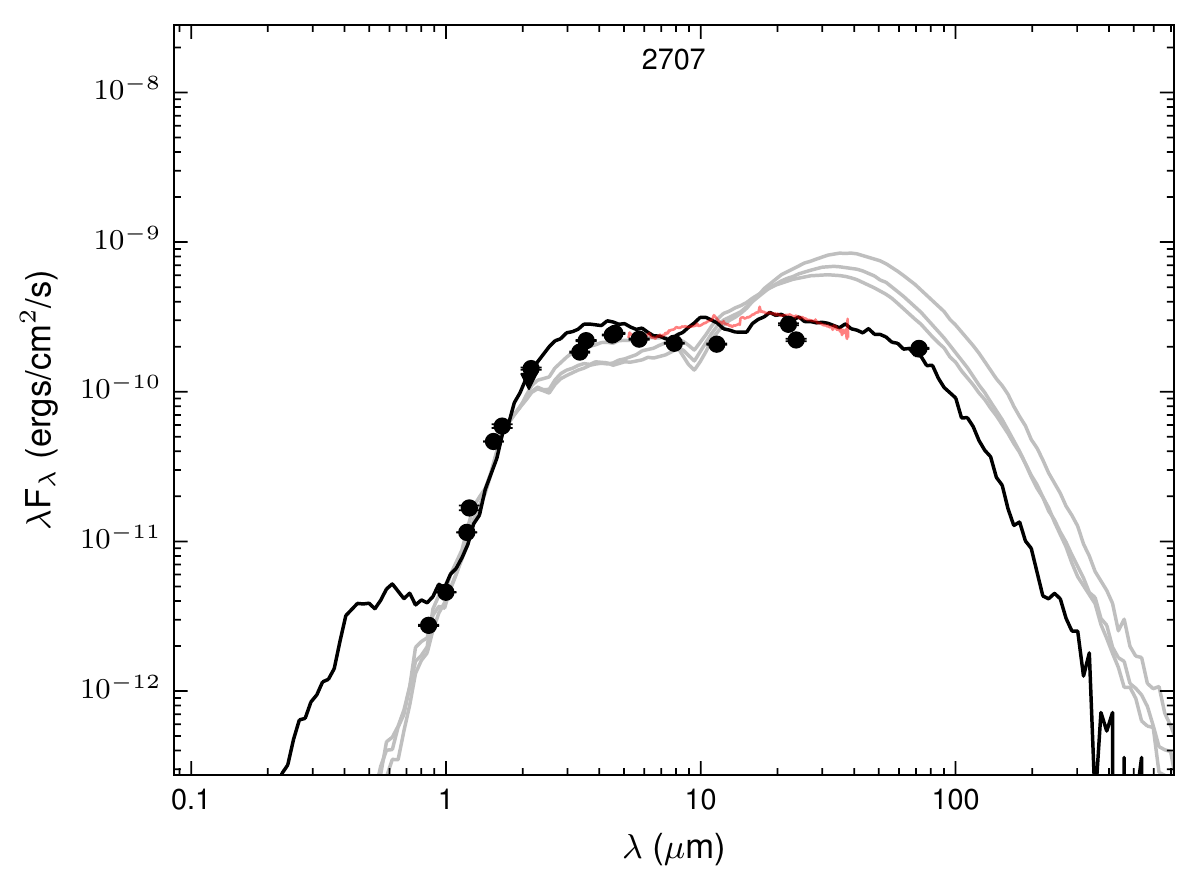} & \ppcell{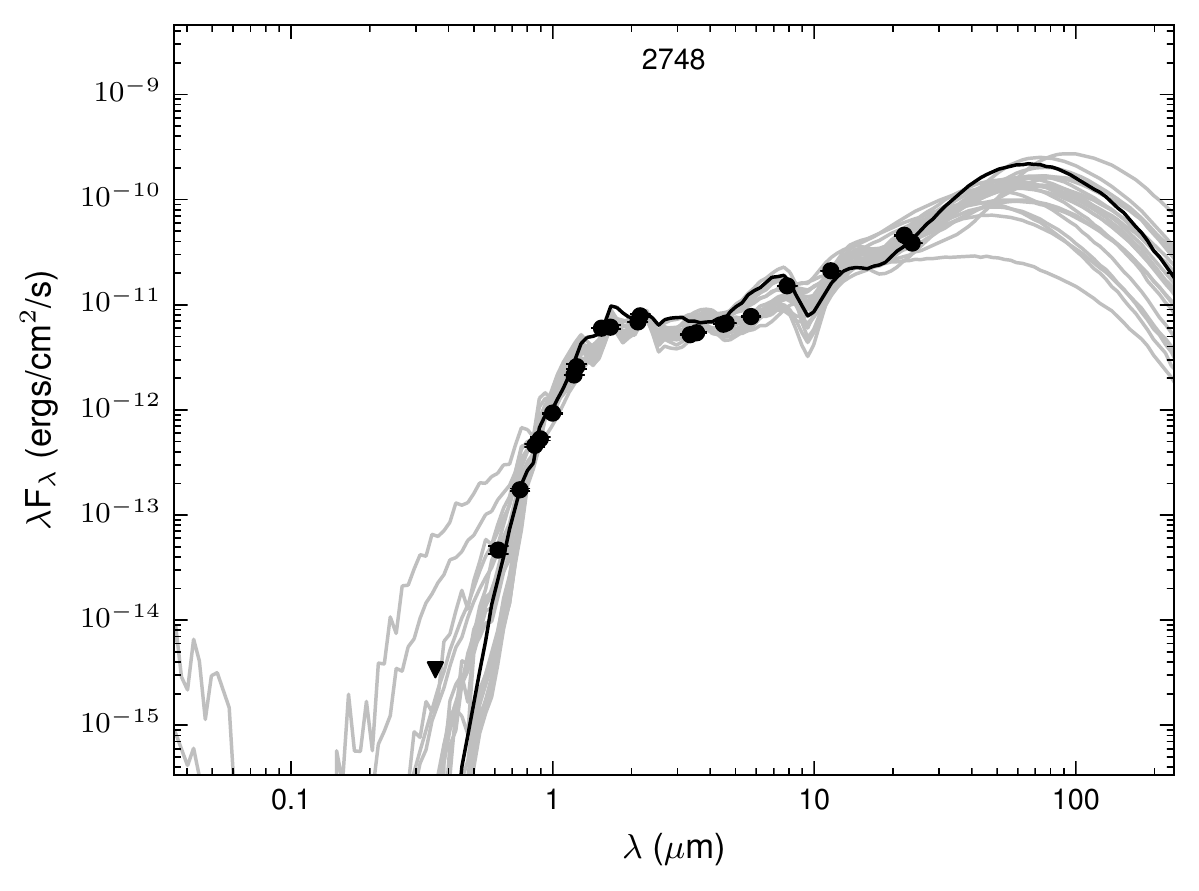} 
\end{longtable}
\end{center}

\end{document}